\documentclass[journal]{IEEEtran}
\IEEEoverridecommandlockouts
\usepackage{cite}
\usepackage{amsmath,amssymb,amsfonts}
\usepackage{graphicx}
\usepackage{textcomp}
\usepackage{xcolor}
\def\BibTeX{{\rm B\kern-.05em{\sc i\kern-.025em b}\kern-.08em
    T\kern-.1667em\lower.7ex\hbox{E}\kern-.125emX}}

\definecolor{blue}{rgb}{0,0,0}

\usepackage{glossaries}
\usepackage{subcaption}
\usepackage{gensymb}
\usepackage{multirow}
\usepackage{tablefootnote}
\usepackage{url}
\usepackage{amssymb}
\usepackage{algorithm}
\usepackage{algpseudocode}
\usepackage{setspace}
\usepackage{graphicx, caption, subcaption}
\usepackage{array}
\usepackage[subtle]{savetrees}

\newacronym{dtn}{DTN}{Delay Tolerant Network}
\newacronym{icn}{ICN}{Intermittently Connected Network}
\newacronym{bp}{BP}{Bundle Protocol}
\newacronym{gs}{GS}{Ground Station}
\newacronym{cp}{CP}{Contact Plan}
\newacronym{cgr}{CGR}{Contact Graph Routing}
\newacronym{ietf}{IETF}{Internet Engineering Task Force}
\newacronym{ccsds}{CCSDS}{Consultative Committee for Space Data Systems}
\newacronym{bps}{bps}{bits per second}
\newacronym{eto}{ETO}{Earliest Transmission Opportunity}
\newacronym{bdt}{BDT}{Best-case Delivery Time}
\newacronym{sr}{SR}{Source Routing}
\newacronym{tdm}{TDM}{Time Division Multiplexing}
\newacronym{raan}{RAAN}{Right Ascension of the Ascending Node}
\newacronym{ltp}{LTP}{Licklider Transmission Protocol}
\newacronym{ion}{ION}{Interplanetary Overlay Network}
\newacronym{crp}{CRP}{Contact Review Procedure}
\newacronym{csp}{CSP}{Contact Selection Procedure}
\newacronym{ip}{IP}{Internet Protocol}
\newacronym{gnn}{GNN}{Graph Neural Network}
\newacronym{drl}{DRL}{Deep Reinforcement Learning}
\newacronym{ml}{ML}{Machine Learning}

\usepackage[colorlinks=true,linkcolor=blue,citecolor=blue,urlcolor=blue]{hyperref}

\begin{document}

\title{Improved Contact Graph Routing in Delay Tolerant Networks with Capacity and Buffer Constraints
	\\
}

\author{Tania Alhajj and Vincent Corlay
\thanks{
			The authors are with Mitsubishi Electric Research and Development Centre Europe, 35700 Rennes, France (e-mail: v.corlay@fr.merce.mee.com).
		} 
}

\maketitle

\begin{abstract}
Satellite communications present challenging characteristics. 
Continuous end-to-end connectivity may not be available due to the large distances between satellites. 
Moreover, resources such as link capacity and buffer memory may be limited. 
Routing in satellite networks is therefore both complex and crucial to avoid packet losses and long delays. 
The Delay Tolerant Network (DTN) paradigm has emerged as an efficient solution for managing these challenging networks.
Contact Graph Routing (CGR), a deterministic routing algorithm, is one of the most popular DTN algorithms. 
CGR is compatible with the ``store, carry, and forward" principle, whereby a node receives a message and stores it in its buffer until a transmission opportunity becomes available.
However, CGR relies on simplified models to incorporate potential constraints in the route search. For instance, the linear volume assumption is often used to consider capacity constraints. 
Moreover, capacity management and buffer management are mostly performed during the forwarding phase, once an issue has occurred. 
\textcolor{blue}{These reactive management techniques cause many collisions and increase the average delivery time.}
In this paper, we propose taking measures before or during the route search in order to find routes that respect both contact-capacity and node-buffer limits. 
We introduce the \textit{contact splitting} and \textit{edge pruning} operations to effectively account for the routing constraints.
This ensures that CGR outputs the optimal solution \textcolor{blue}{in terms of delivery time} among the subset of valid solutions. \textcolor{blue}{The problem is formalized as the Feasible Earliest-Arrival Path with Capacity and Buffer constraints (FEAP-CB) and optimality is proved.}
The proposed approach can also be used to book resources to be used in case of issues during the forwarding phase.
\end{abstract}


\section{Introduction}

Satellite networks are becoming increasingly important for many applications such as global coverage \cite{SatTerNetSurvey2020} and scientific missions in space \cite{CommsuppArtemis21}. 
These networks often face challenges such as long delays, limited connectivity, dynamic topology, and limited resources. Routing algorithms play an important role in this category of networks to ensure efficient data transmission.

As a first networking solution for satellite systems, traditional \gls{ip}-based routing algorithms have been adapted for satellite scenarios where the network topology can be assumed stable and where the connectivity can be considered continuous. 
These \gls{ip}-based algorithms require explicit topology-update notifications \cite{routinginNGsats2007}\cite{IPSat2005}.
In case of frequent network-topology changes, this approach therefore results in significant overhead which can drain the limited storage and processing resources of satellites. Moreover, these \gls{ip}-based algorithms are designed for scenarios where end-to-end connectivity is available. However, satellite networks are often characterized by intermittent connectivity. This intermittent nature induces multiple discontinuous episodes of communication opportunities between pairs of nodes (so-called contacts, see \textcolor{blue}{Section~\ref{sec: keyNotions}}). Limited effectiveness is therefore observed in non-dense satellite scenarios, such as the initial-stage deployment of a constellation.

Since traditional Internet protocols are not suited to the mentioned challenges, the need for new protocols arose. 
The \gls{dtn} \cite{DTNsurveyKhabbaz2012} architecture with a new protocol layer has been proposed by the \gls{ietf} to handle these issues. 
The \gls{bp} \cite{BPspecCCSDS2015}, considered by the \gls{ccsds}, is an instance of such protocols. 
One of the main advantages of a \gls{dtn} architecture is the ability to ``store, carry, and forward" data when immediate transmission is not possible.  
\glspl{dtn} have become essential in modern space communications and are being used in recent projects \cite{nasaHighRateDelay, kplo2022}.

In addition to suitable network capabilities via the above-mentioned protocols, efficient routing algorithms are also needed.
Routing algorithms in \glspl{dtn} can be classified into three categories depending on the availability of network information. 
They can be opportunistic \cite{SprayAWSpyropoulos2005, Vahdat2009EpidemicRF,EnhancedMsgRep2023}, probabilistic \cite{MaxPropBurgess2006}, or deterministic \cite{CGRtutoFRAIRE2020}, depending on whether the contacts are unpredictable, determined with a certain probability, or completely predetermined, respectively.

Opportunistic algorithms rely on message replication, aiming to ensure that at least one copy reaches the destination. 
They are often categorized as epidemic algorithms \cite{Vahdat2009EpidemicRF}. 
These algorithms are highly efficient when energy, buffer space, and link capacity are high or unlimited, as they achieve high delivery ratio and low latency. 
However, such ideal conditions are unrealistic, especially in satellite communication systems where capacity and buffer space are limited and must be used efficiently. 
To address this challenge, algorithms that limit the number of copies have been proposed \cite{SprayAWSpyropoulos2005,EnergyEpidemic2013,EnhancedMsgRep2023}. In the latter reference, the number of copies generated by each node is adjusted based on its resource availability. 
Nevertheless, even with limited copies, resource wastage persists, as the forwarding process remains opportunistic. 
In satellite networks, this wastage can be significantly minimized by leveraging the predictable nature of satellite motion. 

Probabilistic algorithms use statistical models and historical data to predict future network information (i.e., node encounters or contacts) \cite{ProphetSurvey2017}. However, such algorithms are not useful for most satellite networks where node connectivity patterns can be pre-determined. 
 
Deterministic algorithms use comprehensive and pre-computed network information, such as satellite connectivity windows (i.e., contacts) for efficient routing. Unlike opportunistic methods, deterministic algorithms avoid unnecessary transmissions by calculating optimal routes in advance, ensuring that packets (also called bundles, the data units of the \gls{bp}) are forwarded only when a path to the destination is available. 
This approach aligns well with the predictable topology of satellite networks, where orbital mechanics allow precise predictions of link availability.

In this paper, we focus on space communications, where natural and artificial satellites have deterministic motions. Thus, contacts can be predetermined, making deterministic routing algorithms suitable.
The widely used deterministic routing algorithm in \glspl{dtn} is \gls{cgr} \cite{CGRtutoFRAIRE2020}. \gls{cgr} is well suited to satellite systems with limited resources. 
By prioritizing routing based on link availability and predetermined contact duration, the algorithm minimize unnecessary transmissions and optimize delivery time. 
The reader is invited to consult Section~\ref{sub_sec_cgr_histo} for a review of recent improvements of CGR.

Moreover, in addition to the delivery-time objective, optimizing the delivery rate and the resource consumption should also be considered. 
For instance, the use of transmission resources must be carefully optimized because of their limited availability. 
Additionally, as data units are often held in the node buffers for long periods, optimizing storage resources to avoid congestion and loss is crucial.
Nevertheless, the current version of \gls{cgr} and the related latest research do not fully consider or fail to optimize contact capacity and buffer occupancy. Instead, they only provide local and reactive management during the forwarding phase. This leads to potential resource waste, delays, errors, and more generally sub-optimal solutions which may be harmful especially in a resource-constrained environment \textcolor{blue}{(see also Section~\ref{sec: capa challenges} for more details on this aspect)}.  It is therefore important to explore alternative approaches.

One modern application where capacity and buffer management may be critical is cislunar communications. With the advent of the Artemis Program \cite{CommsuppArtemis21}, many projects in the lunar environment have arisen, triggering a high need for communication resources \cite{LunConstMelco22,NASAarch2023}. 
Another application is low-earth orbit satellite constellations, which are expected to be utilized extensively in the future. 
\textcolor{blue}{
Moreover, the IOAG Mars $\&$ Beyond Communications (MBC) architecture explicitly plans the infusion of DTN-based network management to orchestrate relay orbiters, surface hubs, and Earth stations.\cite{tai_lanucara_spaceops_2023}\cite{ioag_mbc_2025}. 
}

In this paper, we highlight the need to enhance the DTN routing algorithms, mainly \gls{cgr}, to consider capacity and buffer management before or during the route search, bridging a critical gap in the current literature.



\textbf{Main contributions.}
We provide a routing solution based on \gls{cgr} with a global and proactive congestion control. 
We propose to ensure contact-capacity management and buffer-limit management before or during the route search such that the obtained route is optimal and respects these two constraints. \textcolor{blue}{Optimality means achieving the lowest possible delivery time. The formal optimization problem (FEAP-CB) is given in Section~\ref{sec:challenges}, and the optimality proof in Section~\ref{sec_opti}}.
The objective is achieved as follows.
\begin{itemize}
\item We propose managing the contact capacity before the route search by modifying the network information, such as the provided contact lists, using contact splitting. Contact splitting removes the part of the contacts used by a routed bundle.
\item We propose to track node buffer occupancy as function of time using forecast buffer tables. The buffer occupancy is then managed by:
\begin{itemize}
\item Temporarily modifying the network information, also via contact splitting, before routing a specific bundle to avoid buffer overflows.
\item Temporarily forbidding some edges between contacts (i.e., a given contact succession) during the route search for a specific bundle to avoid buffer overflows. 
\end{itemize}
\end{itemize}
The modified network information enables the removal of routes that do not respect the constraints (and only these routes). Subsequently, the route-search algorithm, such as \gls{cgr}, can find the optimal solution among the subset of valid solutions.  \textcolor{blue}{It means that the algorithm will automatically find the optimal storage location of bundles  (from delivery time perspective) considering the forecasted buffer status.}

The proposed approach can also be used to reserve resources to be utilized in case of failure events during the forwarding phase. 
These resources, referred to as safety margin in the paper, are to be used by intermediate nodes having limited information on the network state. 

\textcolor{blue}{
\textbf{Assumptions in this paper.}
This study proposes a solution in an ideal scenario. We have the main three assumptions:
\begin{itemize}
\item We assume instantaneous sharing of the network information among source routing nodes, perfect topology knowledge, no unexpected disruptions.
\item We do sequential per packet optimization not full traffic optimization.
\item Complexity is not the main concern.
\end{itemize}
Under these assumptions, we show that the proposed solution avoids all routing errors, both theoretically and with simulations.
We acknowledge that these assumptions are not realistic. This work can therefore be understood as a theoretical foundation under ideal assumptions and may serve as a benchmark in the future for suboptimal solutions operating with partial knowledge, such as \gls{ml}-based approaches.}

\textcolor{blue}{Nevertheless, note that \cite[Sec. IV]{deJonckere2024ContactSegmentation} also investigates the notion of contact segments (conceptually close to contact splitting, proposed at the same period) but considers the complementary simulations in a distributed setting with local, unsynchronized volume views (i.e., without assuming a globally shared, up‑to‑date CP volume state). Simulations show the advantage also in this case, see Section~\ref{sec_contact_segment} for more details.
}

\textbf{Structure of the paper. }The paper is structured as follows. 
In Section \ref{sec: background}, we first introduce key terms used in the paper and we provide a review of \gls{cgr}.
In Section~\ref{sec:challenges}, we focus on existing capacity and buffer management challenges in \glspl{dtn} to introduce our motivations \textcolor{blue}{and we present the mathematical problem formulation (FEAP-CB)}.
\textcolor{blue}{The proposed solution is described in Section~\ref{sec: global management}, which also includes a proof of optimality of the approach.} The source-routing \gls{cgr} benchmark is detailed in Section \ref{sec: benchmark}. Simulation assumptions and results comparing our method with the benchmark are given in Sections \ref{sec: setup} and \ref{sec: results}, respectively. \textcolor{blue}{Symbols used in the paper are summarized in the Appendix (Table~\ref{tb: notation}).}

\section{Background and motivations}
\label{sec: background}

\subsection{Key notions}
\label{sec: keyNotions}

We first define some key terms used in this paper.

\begin{figure}[!t]
	\centering
	\includegraphics[scale=0.35]{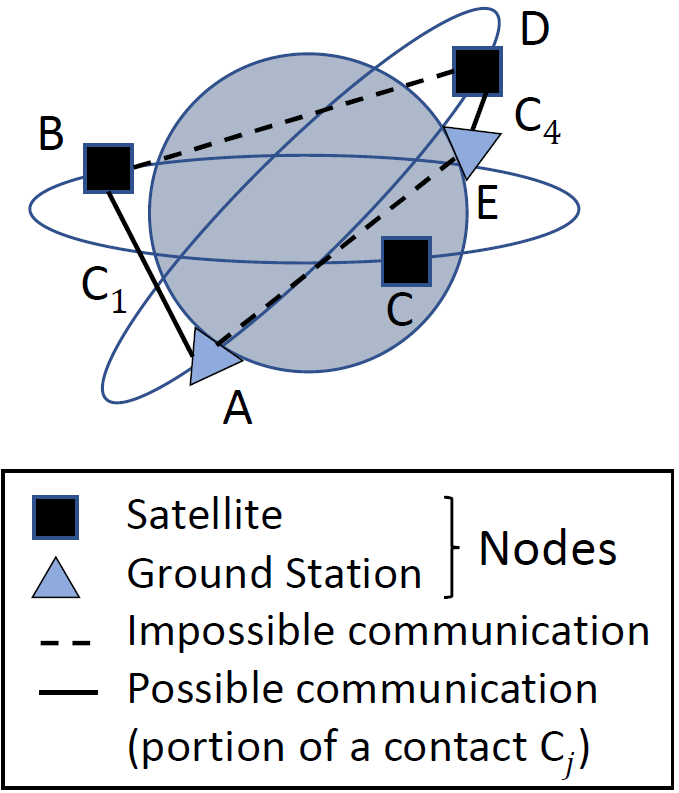}
	\caption{Snapshot of a satellite communication scenario at time $t$ (e.g., $ t_{1,1}\leq t \leq  t_{2,1}$ and $ t_{1,4}\leq t \leq  t_{2,4}$, see Table \ref{tb: CP}).}
	\label{fig: SatConst}
\end{figure}

\begin{itemize}
	\item Node: In a satellite communication scenario, a \emph{node} can be a satellite or a \gls{gs} as illustrated in Figure \ref{fig: SatConst}. \textcolor{blue}{The set of all nodes is $\mathcal{N}$. In the text we denote a node by $n$ or by context-specific letters ($s$, $d$ for source and destination of a bundle, $u_i$, $v_i$ for sender and receiver of contact $\text{C}_i$). In the figures we use capital letters ($X$, $A$, $B$, etc.) for clarity.}
	
	\item Contact: A \emph{contact} $\text{C}_j$ is defined by a communication opportunity between two nodes. It is characterized by sender and receiver nodes, a start time $t_{1,j}$, an end time $t_{2,j}$, and a \emph{data rate/capacity} $R_j$ in \gls{bps}. The \emph{volume} of a contact is its rate multiplied by its duration. 
\textcolor{blue}{Each contact $\text{C}_i$ is therefore a tuple
\begin{align}
	\begin{split}
&\text{C}_i=(u_i,\,v_i,\,t_{1,i},\,t_{2,i},\,R_i,\,\delta_i),\quad\\
& t_{1,i}<t_{2,i},\ \ R_i>0,\ \ \delta_i\ge 0,
	\end{split}
\end{align}
where $\delta_i\ge 0$ is the one-way propagation delay.}
	
	\item \gls{cp}: A \emph{\gls{cp}} is the list of contacts for a time period. It reflects the evolution of the network topology over time. An example is provided in Table~\ref{tb: CP}.
\textcolor{blue}{The set of contacts is either denoted by $V_c^0$ or simply CP.}
	
	\item Bundle: A \emph{bundle} is the data unit encoded by the \gls{bp}. A bundle $i$ is characterized by its size $S_i$ in bits.
\textcolor{blue}{A bundle of size $S>0$ is released at source $s\in\mathcal{N}$ at time $t_0$ and must reach
destination $d\in\mathcal{N}$.}
	\item Edge: An \emph{edge} $(i\to j)$ exists between two contacts when the receiver of the first contact $i$ matches the sender of the second contact $j$ ($v_i=u_j$).
\textcolor{blue}{$E_c^0=\{(i\to j): v_i=u_j\}$ is the set of such edges, for all contacts in $V_c^0$.}\\
\textcolor{blue}{For a given bundle of size $S$, the \emph{feasible window} of contact $\text{C}_i$ is $W_i=[t_{1,i},\,t_{2,i}-S/R_i]$: the set of start times at which a transmission of $S$ bits on $\text{C}_i$ finishes by $t_{2,i}$. Thus each transmission must finish before the contact ends, and each next hop must start no earlier than the arrival $f_i=\tau_i+S/R_i+\delta_i$ from the previous hop. For a given bundle of size $S$, only a subset of $E_c^0$ that satisfy these timing constraints is actually usable\footnote{\textcolor{blue}{An equivalent modeling is to enforce this condition and the schedulability directly in the contact graph by reducing $E_c^0$.}}.}
	\item  \textcolor{blue}{Buffer capacity.
For each node $n$, let $B_n(t)$ denote the (full) \emph{buffer capacity} (in bits) available at time $t$. In this paper, the buffer capacity is constant over time, so $B_n(t)=B_\text{max}$ for all $t$.
Let $Q_n(t)$ be the booked buffer load due to previously scheduled bundles ($Q_n(t)=0$ if no prior bookings).}
\textcolor{blue}{We define the residual buffer capacity as
\[
\bar B_n(t)=\max\{\,0,\ B_n(t)-Q_n(t)\,\}.
\]}

\item \textcolor{blue}{Connectivity-only contact graph $(V_c^0,E_c^0)$: We denote by $(V_c^0,E_c^0)$ the \emph{connectivity-only} contact graph before any splitting: vertices are the contacts in the \gls{cp}, and edges are as in the ones described in the paragraph above.}
\item \textcolor{blue}{Feasible contact graph $(V_c,E_c)$: Contact splitting (see Section~\ref{sec: ct splitting intro}) turns $V_c^0$ into the vertex set $V_c$ of split contacts. On $V_c$, the connectivity-only edge set is again denoted $E_c^0$. The \emph{feasible} contact graph $(V_c,E_c)$ is obtained by removing additional edges from $E_c^0$ via \emph{edge pruning} (see Section~\ref{sec_edge_prunning_prez}), so $E_c\subseteq E_c^0$. How $(V_c,E_c)$ is built is detailed in Section~\ref{sec: global management}.}
\item \textcolor{blue}{Route (path) and schedulability:
A \emph{route} (or \emph{path}) is a sequence of contacts $P=(\text{C}_1,\ldots,\text{C}_k)$ used to deliver a bundle from source $s$ to destination $d$, with the basic connectivity constraint that the receiver of $\text{C}_i$ equals the sender of $\text{C}_{i+1}$:
\[
u_1=s,\quad v_k=d,\quad v_i=u_{i+1}\ \ (i=1,\ldots,k-1).
\]
In the connectivity-only contact graph $(V_c^0,E_c^0)$, this means $\text{C}_i\in V_c^0$ and $(\text{C}_i,\text{C}_{i+1})\in E_c^0$ for all $i$, so $P$ is a path in $(V_c^0,E_c^0)$.}

\textcolor{blue}{When considering transmission timing for a given bundle, a path $P=(\text{C}_1,\ldots,\text{C}_k)$ is \emph{schedulable} if it satisfies the timing constraints (feasible windows and precedence) introduced in the Edge paragraph above. This schedulability is enforced in the path selected by \gls{cgr} when the route search runs on the contact graph: the (time-dependent) Dijkstra search explores only contact successions that respect these constraints.}
\item \textcolor{blue}{Feasible route: A \emph{feasible route} is a path (in the connectivity graph $(V_c^0,E_c^0)$) that is schedulable, contact-capacity feasible, and buffer feasible (see Section~\ref{sec_math_sub}). The set of feasible routes for a bundle of size $S$ and release time $t_0$ is denoted $\mathcal{P}_f(S,t_0)$. We will show (Lemma 3, Section~\ref{sec_opti}) that $\mathcal{P}_f(S,t_0)$ coincides with the set of schedulable paths in the feasible graph $(V_c,E_c)$.}
\end{itemize}

\subsection{Contact graph routing (history)}
\label{sub_sec_cgr_histo}

Since its invention, \gls{cgr} has gained attention in the research community and several improvements have been proposed. 
Using Dijkstra's algorithm for the route search and choosing the \gls{bdt} as objective were presented in \cite{eCGRSegui2011}. Then, the \gls{eto} was introduced to compute the time at which a bundle is transmitted through the contact \cite{ETOBezirgiannidis2013}. The \gls{eto} takes into account the time a bundle has to wait before being transmitted because of other scheduled bundle transmissions on the same contact. Later, a complementary overbooking management was added to the \gls{eto} to improve routing decisions \cite{EnhBezirgiannidis2014}. Nevertheless, these management \textcolor{blue}{techniques} are performed during the forwarding phase. Furthermore, a route list-management technique using Yen's algorithm was proposed to further enhance the standard \gls{cgr} algorithm \cite{FRAIREYenCGR2018}. Note however that this route management technique increases processing complexity during scheduling \cite{YenTract2023}.
These enhancements among others result in the currently used version of \gls{cgr} \cite{CGRtutoFRAIRE2020}. 

\textcolor{blue}{\subsection{Recent trends in DTN routing and further elements on the SoA}}

Research continues to improve the latest version of \gls{cgr}, focusing on the optimization of limited resources including energy \cite{EnergyCGR2020}, computation \cite{cacheCGR2014, SRMarchese2017}, heterogeneous traffic and buffer-related aspects \cite{HetTrafficDTN2023}, and link capacity \cite{OptCGR2020}.

\subsubsection{\textcolor{blue}{Contact segmentation in \cite{deJonckere2024ContactSegmentation}}}\label{sec_contact_segment}
\textcolor{blue}{Contact segmentation was proposed independently around the same period as the first version of this paper \cite[Sec. IV]{deJonckere2024ContactSegmentation}. It follows the same idea as contact splitting: instead of splitting the contacts, that approach adds a new parameter to the contacts which describes the available \emph{segments} of the contact. That work however considers simulations for the distributed case with local, unsynchronized volume views (i.e., without assuming a globally shared, up‑to‑date CP volume state). In \cite{deJonckere2024ContactSegmentation}, the capability of the node is implemented using a new library \cite{deJonckere2025ASABR}.}

\textcolor{blue}{The following improvement compared to the benchmark is summarized in the conclusion of \cite{deJonckere2024ContactSegmentation}: contact segementation delivers up to a 10\,\% improvement in delivery ratios, reductions in delivery delay of up to 11\,\% and up to a 16\,\% decrease in hop counts.}

\textcolor{blue}{\subsubsection{Machine-learning routing.}
Recent works apply \glspl{gnn} and \gls{drl} to LEO/DTN routing to adapt under dynamic environments, potentially with partial network knowledge and unexpected events. This is promising for \emph{traffic-level} optimization where load, not just individual bundles, must be steered and complexity is an issue.
However, these methods generally \emph{lack shortest-path optimality guarantees} and usually handle contact-capacity and buffer limits via rewards or post-hoc checks rather than \emph{hard feasibility at selection time} \cite{Jiang2024GNNRoutingSurvey,Zheng2024DynamicGNNsurveyArxiv,LozanoCuadra2025Xplore}. Therefore, current research investigates the trade-off between the theoretical guarantees given by CGR under simplified assumptions and the flexibility brought by \gls{ml}.
}

\textcolor{blue}{We review some existing works.
GAUSS \cite{Olmedo2023GAUSS} learns to accelerate route selection in scheduled DTNs, reporting up to $2\times$ compute reduction and $3\times$ lower variability versus CGR-style search. However, Section~4.1.2 of that work explicitly notes that this better computational profile comes at the expense of decreased efficacy, with degraded delivery time compared to standard CGR.}\\
\textcolor{blue}{Under congestion or schedule errors, learned policies can rebalance load and improve throughput/delay, making \gls{ml} a practical \emph{fallback} or \emph{online refinement} atop deterministic baselines \cite{Han2024DMR,Shi2024GraphSAGEDQN}.
Other works learn next-hop or path-selection policies and show gains in simulations, e.g., multi-agent \gls{drl} adapts under load. \gls{drl}+\gls{gnn} combinations improve aggregate metrics in mega-constellations \cite{LozanoCuadra2024MADRL,Han2024DMR,Shi2024GraphSAGEDQN}. GraphSAGE+DQN and graph attention handle topology changes \cite{Shi2024GraphSAGEDQN,GATLEO2025}.}

\textcolor{blue}{\subsubsection{Buffer-aware routing and resource reservation in DTNs.}
Many opportunistic DTN schemes have long intertwined routing with \emph{buffer management}. As mentioned in the introduction, opportunistic refers to networks with stochastic contacts and typically rely on replication/priority and local queue control to mitigate uncertainty \cite{Jain2014BufferSurvey,Singh2023BufferSurvey,RFC6693}. This line of work also considers the dropping policy at the buffer node, which is beyond the scope of this paper.}

\textcolor{blue}{Within this opportunistic family, \emph{MaxProp} orders both transmissions and \emph{drops} (explicit buffer eviction policy) using path-likelihood estimates and global ACKs \cite{Burgess2006MaxProp}. \emph{RAPID} treats routing as a \emph{resource allocation} problem, replicating bundles in decreasing marginal utility to optimize metrics such as mean delay or deadline-meet ratio \cite{Balasubramanian2007RAPID}. \emph{PRoPHET} prunes epidemic flooding via delivery predictabilities. Variants such as PRoPHETv2 add energy/buffer awareness in forwarding and queue control \cite{Lindgren2011PRoPHETv2}. Constrained \emph{Spray-and-Wait} methods cap replication and incorporate congestion control or dynamic copy allocation to reduce buffer pressure \cite{Oham2015CASaW,Zhang2012DSW,Li2015SWSB}. Beyond replication caps, \emph{ORWAR} is explicitly \emph{resource-aware}: it estimates contact windows to avoid partial transmissions and orders replication/drops by message utility (buffer-, time-, and contact-aware scheduling) \cite{Sandulescu2008ORWAR}. This is closer to the approach proposed in this paper, as our routing algorithm considers forecasted buffer resources when choosing the route.
Recent surveys emphasize that these approaches largely enforce \emph{soft} feasibility (via prioritization, dropping, replication limits), whereas our CGR formulation guarantees feasibility during \emph{route computation} in scheduled DTNs \cite{Jain2014BufferSurvey,Singh2023BufferSurvey,SABRspecCCSDS2019}.}

\subsection{Contact graph routing (implementation)}
\label{sec: CGR}

\gls{cgr} is divided into three main phases \cite{SABRspecCCSDS2019,CGRtutoFRAIRE2020}. 
First, the \gls{cp}, for example constructed at a mission control unit, is forwarded to all nodes in the network. 
Second, in the forwarding (i.e., transmission) phase, each node receiving a bundle performs the route search using Dijkstra's algorithm to determine the next hop. 
Third, checks are performed at each intermediate node before transmission to verify that system constraints are respected.

	\subsubsection{\gls{cp} construction} In a deterministic network case, the \gls{cp} is constructed at the source nodes using the network information and transmitted to all intermediate nodes for the route-search phase. 
	
		\subsubsection{Route search} 
Route search is performed using an adapted version of Dijkstra's algorithm \cite{Dijkstra1959}. This adaptation considers the discontinuities of \glspl{icn} by using a contact-graph representation of the network rather than a node-graph representation. \textcolor{blue}{In this standard implementation, the route search runs on the connectivity-only contact graph $(V_c^0,E_c^0)$. In the proposed algorithm (Section~\ref{sec: global management}), it runs on the feasible graph $(V_c,E_c)$.} The output of a Dijkstra search is the route, consisting of a succession of contacts from the \gls{cp}, connecting the source to the destination with the lowest \gls{bdt}.

The detailed implementation can be found in \cite{CGRtutoFRAIRE2020}. Here, we explain the aspects of the route search that are essential for understanding our contribution.
The idea is to explore the available contacts in the \gls{cp} to determine the shortest route. 
The exploration of the \gls{cp} is performed by analysing successively selected contacts denoted by $\text{C}_s$. 
For the first iteration, the selected contact $\text{C}_s$ is a root contact, which is virtual and whose sender and receiver are the bundle's source node.
At each iteration, a new contact $\text{C}_s$ is selected and used in the next iteration. 
Two main procedures are executed at each iteration. The first procedure is the \gls{crp}, used to explore the \gls{cp}. The second one is the \gls{csp}, used to determine the following $\text{C}_s$. 

Algorithm \ref{alg:CRP} presents both the original CRP and our contributions. Our modifications are marked with asterisks (**, lines $2$ and $8-12$) in the algorithm.
In this section, we explain only the original \gls{crp}. The modifications will be explained in Section \ref{sec:  proposedBuffManagement}. 

In the \gls{crp}, the \gls{cp} is explored starting from a selected contact $\text{C}_s$. 
During this procedure, all contacts are considered for exploration. 
In the algorithm, the following characteristics of a contact are updated and used:
\begin{itemize}
\item A successor of a contact (the contact that comes after in a route). \textcolor{blue}{A successor of $\text{C}_s$ is a contact $\text{C}_i$ with $(\text{C}_s,\text{C}_i)\in E_c^0$.}
\item A predecessor of a contact (the contact that comes before in a route).
\item An arrival time $f_i$ at the receiver of the contact. \textcolor{blue}{$f_i=\tau_i+S/R_i+\delta_i$.}
\item Visited nodes, which are the nodes that exist in the route or the succession of contacts that led to this contact.
\item A visited flag, which indicates if this contact has been visited, meaning that it has been previously selected and used to start exploring contacts in the \gls{cp}.
\item A suppressed indicator, which indicates if the use of the contact is forbidden (for security reasons for example).
\end{itemize}
A contact $\text{C}_i \in$ \gls{cp} must meet the following criteria to be explored (otherwise it is ignored) (lines $5-7$). If these criteria are met $\text{C}_i$ will be considered as a potential successor of $\text{C}_s$.  \\
a) $\text{C}_i$ must have the selected contact's receiver as its sender. \textcolor{blue}{(Connectivity: $v_s=u_i$, so $(\text{C}_s,\text{C}_i)\in E_c^0$.)} \\
b) For a bundle of size $S$, $\text{C}_i$ is \textcolor{blue}{schedulable if $f_s \le t_{2,i}-S/R_i$
and $t_{1,i} \le t_{2,i}-S/R_i$, that is, the earliest feasible start
$\tau_i^\star=\max\{t_{1,i},f_s\}$ lies in the feasible window
$W_i=[t_{1,i},\,t_{2,i}-S/R_i]$. For readability,
Algorithm~\ref{alg:CRP} reference the \emph{weaker} condition $t_{2,i}\le f_s$.} \\
 c) $\text{C}_i$ should not have been visited, meaning that it should not have been selected as $\text{C}_s$ in a previous \gls{crp} iteration.\\
  d) The receiver node of $\text{C}_i$ must not be visited i.e., be a node of the contact sequence that led to $\text{C}_i$, to avoid loops. \\
   e) $\text{C}_i$ must not be suppressed (e.g., removed or forbidden). \\
  f) The remaining volume of $\text{C}_i$ must be greater than zero. \textcolor{blue}{(So the contact has capacity for the bundle, and the feasible window $W_i$ is non-empty.)} This means that after allocating $\text{C}_i$ to bundles and deducting bundle size from the volume of $\text{C}_i$, some volume must remain available in $\text{C}_i$ to allow more data to pass through.

Once a contact has passed all checks, it is considered as an explored contact.
The arrival time of an explored contact is computed according to the time taken to reach the receiver of the explored contact following the route that led to it (line $13$). 
If the computed arrival time is less than its last calculated value $f_i$ (in the case where $\text{C}_i$ has already been explored), it is updated (lines $14$ and $15$). 
In addition, the selected contact $\text{C}_s$ becomes the predecessor of $\text{C}_i$ and the visited nodes of $\text{C}_i$ are  updated by adding the visited nodes of the nodes  within the succession of contacts that led to $\text{C}_i$ (lines $16$ and $17$) (to avoid getting back to them, which would induce loops). 

Each time the destination node is reached (i.e., when the receiver of an explored $\text{C}_i$ is the bundle's destination), the arrival time $f_i$ is saved as \gls{bdt} if it is lower than the previously calculated \gls{bdt}. In such cases, $\text{C}_i$ is considered as a final contact $\text{C}_\text{end}$ with its corresponding \gls{bdt} (lines $18-21$). After finishing all contacts in the \gls{cp}, the currently selected contact $\text{C}_s$ is marked as visited to avoid repeating the same work later (line $24$).

\begin{algorithm}
\caption{Updated Contact Review Procedure (\gls{crp})}
\label{alg:CRP}
\begin{spacing}{0.8}
\begin{algorithmic}[1]

\State \textbf{Input}: \gls{cp}, $\text{C}_s$, $\text{C}_\text{end}$, \gls{bdt}
\State \textit{**} \textbf{Additional input}: Bundle b, $T_{1,s} = \{t_{1,o_l}\} \mid o_l \in  \mathcal{O}_{Z_1} $
, where ${Z_1}$ is the receiver of $\text{C}_s$ \textit{**}
\State \textbf{Output}: updated \gls{cp}, $\text{C}_\text{end}$, \gls{bdt}
\For {$\text{C}_i \in$ \gls{cp}} 
\If {sender of $\text{C}_i \neq$ receiver of $\text{C}_s$ \textbf{or} $t_{2,i}\leq f_s $ \textbf{or} $\text{C}_i$ is already visited \textbf{or} receiver of $\text{C}_i$ is a visited node by $\text{C}_s$ \textbf{or} $\text{C}_i$ is suppressed \textbf{or} remaining volume of $\text{C}_i == 0$}
\State ignore $\text{C}_i$
\EndIf
\Statex \textit{** Additional condition}
\State $\triangleright$ \emph{Comment: Compute the first overflow start strictly after $f_s$: successors must start \emph{before} it to avoid buffer overflows.}
\State $a_{\text{next}} \gets \min\{\,t \in T_{1,s}\mid t>f_s\,\}$; \textbf{if} no such $t$ \textbf{then} $a_{\text{next}}\gets +\infty$
\If{$t_{1,i} \ge a_{\text{next}}$}
\State ignore $\text{C}_i$
\EndIf
\Statex \textit{**}
\State compute arrival time $f$ at the receiver of $\text{C}_i$ (when extending from $\text{C}_s$)
\If {$f<f_i$}
\State $f_i = f$
\State predecessor of $\text{C}_i = \text{C}_s$
\State $\text{C}_i$'s visited nodes $= \text{C}_s$'s visited nodes and $\text{C}_i$'s  

 \hspace{7pt} receiver 
\If {$\text{C}_i$'s receiver is bundle's destination 
    \textbf{and}
    
     \hspace{7pt} $f_i < $ \gls{bdt}  }
   \State \gls{bdt} $=f_i$
   \State $\text{C}_\text{end} = \text{C}_i$
\EndIf
\EndIf
\EndFor
\State Consider $\text{C}_s$ as visited

\end{algorithmic}
\end{spacing}
\end{algorithm}

In the \gls{csp}, the best contact among the explored contacts from \gls{crp} is selected based on the shortest arrival time. This contact becomes the new selected contact for the next \gls{crp} iteration. The iterations continue until no additional contacts can be selected for further exploration. The final contact of the route is then the one with the lowest \gls{bdt}.

Once these procedures are complete, the route is constructed in reverse order. Starting from the final contact (which has the lowest \gls{bdt}), the algorithm traces back through the predecessors until reaching the root contact. For each Dijkstra call, the \gls{cp} is reset to its initial state (i.e., as if no contacts had been explored).

Yen's algorithm \cite{FindingKPathsYen1971} enables the determination of the $k$ shortest routes by iteratively exploring paths between a source and a destination. 
It uses Dijkstra's algorithm with the contact-graph representation while avoiding loops and previously discovered paths.

In the standard \gls{cgr} implementation, each node computes a list of the $k$ shortest routes to each node in the network using Yen's algorithm, where $k$ is an input parameter. This is referred to as per-hop routing. When a bundle reaches an intermediate node, this node selects a feasible pre-computed route from its route list to determine the next hop.

			\subsubsection{Validation and forwarding} \label{sec: valid}
In the forwarding phase, the system examines both the remaining contact volume and the buffer queuing time for bundles scheduled on the same contact. 
These validation checks verify the feasibility of previously computed routes and prevent local congestion.

The system determines valid route candidates from the $k$ shortest routes. A valid route must guarantee sufficient contact volume for the bundle's size. Additionally, the system must verify that the bundle can complete transmission, including queuing time, before the contact ends (see \gls{eto} \cite{ETOBezirgiannidis2013}).

The system selects the optimal route from valid candidates to determine the bundle's next hop. After the route selection, the system queues the bundle for transmission and decreases the contact's available volume by the bundle's size.

\subsubsection{Additional comments on \gls{cgr}}
Although widely used, \gls{cgr} is computationally complex. Each node must perform multiple checks per bundle simply to determine the next contact. Since validation occurs after the route search, the system might find no valid solutions among the candidate routes. Such situations often trigger time-consuming iterations between Dijkstra's algorithm and validation checks.

\subsection{Source-routing CGR}

\gls{sr} \cite{SourceRoutBIRRANE2012} offers an effective approach to reduce \gls{cgr}'s complexity. Unlike per-hop routing, \gls{sr} computes the complete route at the source node and encodes it into the bundle's extension block \cite{EBirtf2013}.

When a bundle with a pre-calculated route arrives at an intermediate node, a verification of the encoded route's validity is performed. Upon detecting an invalid route, the intermediate node reverts to traditional \gls{cgr} to compute a new route to the destination. The following two types of \gls{sr} can be implemented.

\subsubsection{Bundle-independent \gls{sr}}
In bundle-independent \gls{sr}, the source node computes routes without considering bundle properties or queuing information. During forwarding, the system performs standard \gls{cgr} checks such as \gls{eto} and volume availability. The system validates the encapsulated route using bundle properties (e.g., size) and local queue states \cite{MSRCaini2020}.

\subsubsection{Bundle-dependent \gls{sr}}
\label{ref_bundle_dep_sr}
Bundle-dependent \gls{sr} utilizes available bundle properties (e.g., size and priority) during initial route computation \cite{CGRtutoFRAIRE2020}. During forwarding, the system evaluates route viability using local queue information.

Our solution extends this approach by incorporating both network information and bundle properties during initial route computation.

\section{Contact-capacity and node-buffer challenges in DTNs: \\\textcolor{blue}{Presentation and problem formulation}}
\label{sec:challenges}
\subsection{Contact-capacity management}
\label{sec: capa challenges}

\begin{figure}[!t]
	\centering
	\includegraphics[scale=0.4]{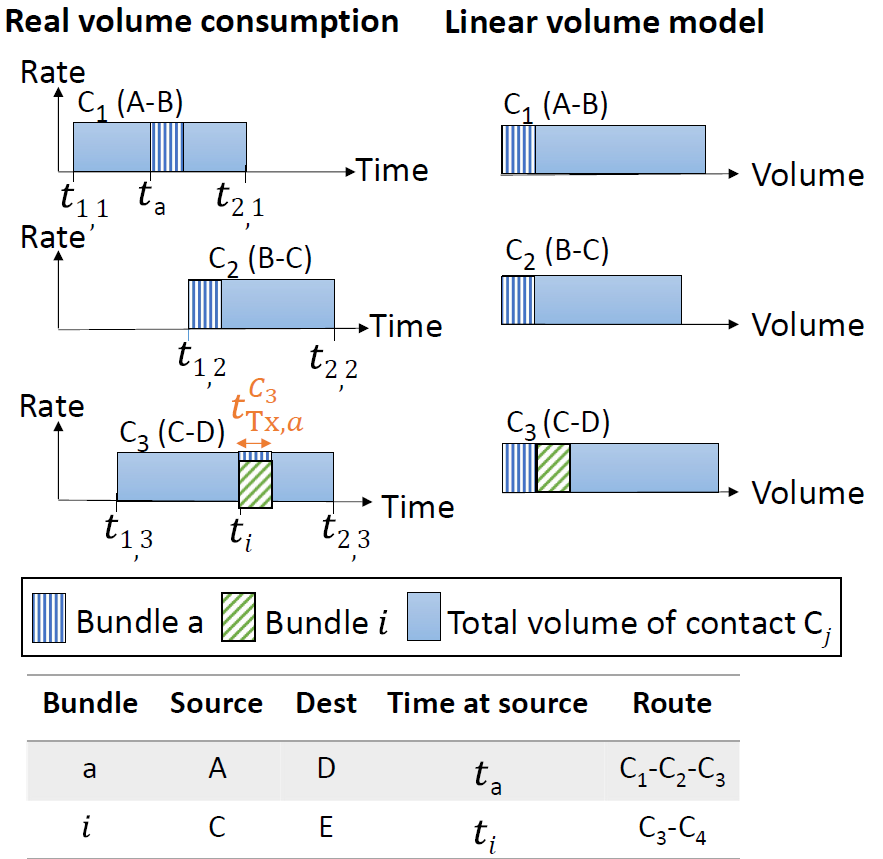}
	\caption{Linear vs real volume consumption.}
	\label{fig: volumeLinVSreal}
\end{figure}

\gls{cgr} manages capacity consumption and availability during the forwarding phase. The route search implements the so-called linear approach that verifies whether available contact volume exceeds the bundle size \cite{CGRtutoFRAIRE2020}. This approach fails to prevent capacity limit violations because it does not track the temporal usage of contacts. The linear volume model is illustrated in Figure \ref{fig: volumeLinVSreal} (right). As shown in the figure, the linear model incorrectly indicates that bundle $i$ (hatched rectangle) arriving at node C at time $t_i$ has immediate access to contact $\text{C}_3$. In contrast, the actual contact volume consumption (Figure~\ref{fig: volumeLinVSreal}, left) reveals that bundle $i$ cannot use contact $\text{C}_3$ immediately because bundle $a$ (striped rectangle) occupies the entire capacity at that time. When traditional \gls{cgr} detects this conflict during forwarding through \gls{eto}, the transmission must be delayed.

Nevertheless, delaying the transmission of bundle $i$ in the forwarding phase may cause multiple problems such as inaccurate estimates of delivery time, the end or unavailability of current or future contacts on the planned route, or reaching the bundle delivery deadline before its delivery.

These collision problems can be better managed, or even avoided, if capacity consumption is modeled differently. In the same example, if the use of $\text{C}_3$ at that precise moment had been known in advance, an alternative route might have been a more effective solution for delivering bundle $i$.

\begin{table}
	\caption{Initial Contact Plan.}
	\label{tb: CP}
\begin{center}
	
\begin{tabular}{|c | c |c |c |c|}
		\hline
		\textbf{Contact} &\textbf{Sender} & \textbf{Receiver} & \textbf{Start time}  & \textbf{End Time}  \\ \hline
		C$_1$ & A & B & $t_{1,1}$ &$t_{2,1}$ \\ \hline  
		C$_2$ & B & C & $t_{1,2}$ &$t_{2,2}$  \\ \hline
		C$_3$ & C & D & $t_{1,3}$ &$t_{2,3}$   \\ \hline
		C$_4$ & D & E  & $t_{1,4}$ &$t_{2,4}$   \\ \hline
	\end{tabular}

\end{center}
\end{table}

\subsection{Buffer management}

Node buffer management is crucial in \glspl{dtn} because bundles may require extended storage before transmission. Buffer capacity constraints further complicate this challenge.

Traditional buffer management strategies react to problems after they occur. When a buffer is full, the system must immediately decide to reroute or drop bundles, as seen in reactive custody transfer \cite{BPspecCCSDS2015}. An alternative approach proposed in \cite{AutonomousCCBurleigh2006} involves refusing custody to preserve node buffer space. However, these strategies only utilize information from neighboring nodes.

An alternative approach consists in implementing global buffer occupancy management during initial route planning. Unlike contact-level volume and capacity management, buffer occupation requires node-level management. Multiple bundles using different contacts can affect the same node's buffer capacity. Consequently, this characteristic demands careful consideration when working with contact-graph network representations.

\textcolor{blue}{\subsection{Mathematical problem formulation}\label{sec_math_sub}}
\textcolor{blue}{We first formalize the constraints to be respected and then state the optimization problem.
Given a path  $P=(\text{C}_1,\ldots,\text{C}_k)$, with a transmission start time $\tau_i$ on contacts $\text{C}_i$, a given bundle size $S>0$, and release time $t_0$ (and source $s$, destination $d$): }

\textcolor{blue}{\subsubsection{Contact-capacity feasibility}
$P$ is \emph{contact-capacity feasible} if it does not use any portion of a contact that is already allocated to another bundle (i.e., each contact used has a sufficient unallocated portion for the bundle).}

\subsubsection{\textcolor{blue}{Buffer feasibility} }
\textcolor{blue}{Given $P$, the bundle is stored at node $v_i$ during the time interval
\begin{align}
\begin{split}
&I_{v_i}=[\,\tau_i+S/R_i+\delta_i,\ \tau_{i+1}\,]\ \ (i=1,\ldots,k-1).
\end{split}
\end{align}
\emph{Buffer feasibility} requires (assuming no storage issue at the source node $s$ and at the destination node $d$)
\begin{align}
\begin{split}
&S \le \bar B_{v_i}(t) \quad\Leftrightarrow\quad
S \le \inf_{t\in I_{v_i}}\bar B_{v_i}(t), \\
& \forall t\in I_{v_i}\ (i=1,\ldots,k-1).
\end{split}
\end{align}
}

\textcolor{blue}{\subsubsection{Feasible route set}
A \emph{feasible route} is a path $P$ in the connectivity graph $(V_c^0,E_c^0)$ that is (i) \emph{schedulable}, (ii) \emph{contact-capacity feasible}, and (iii) \emph{buffer feasible}. We denote the set of feasible routes by $\mathcal{P}_f(S,t_0)$:}
\textcolor{blue}{%
\begin{align}	\begin{split}
&\mathcal{P}_f(S,t_0)= \Big\{\ P=(\text{C}_1,\ldots,\text{C}_k):\ P \text{ path in } (V_c^0,E_c^0) and,\\
& P \text{ schedulable, contact-capacity feasible,} \text{ and buffer feasible}\ \Big\}.
	\end{split}
\end{align}
}

\textcolor{blue}{
We denote the arrival time at destination as
\begin{align}
A(P)=\tau_k + S/R_k + \delta_k=f_{i_k}.
\end{align}
}

\textcolor{blue}{With nonnegative traversal times ($S/R_i+\delta_i\ge 0$) and waiting allowed ($\tau_{i_{m+1}}\ge f_{i_m}$), the time-dependent graph is \emph{FIFO}: on every (feasible) edge, a later departure cannot yield an earlier arrival at the receiver. We introduce this property so that the time-dependent shortest-path (TDSP) framework of \cite{Dean2004TDSP} applies when establishing optimality (Section~\ref{sec_opti}).
}

\textcolor{blue}{\subsubsection{Problem to solve: Feasible Earliest-Arrival Path with Capacity and Buffer constraints (FEAP-CB)}
For a given bundle, the optimization problem is to find $P^*$ minimizing the arrival time among feasible routes $\mathcal{P}_f(S,t_0)$:
\begin{align}
P^*  =  \operatorname*{arg\,min}\{\, A(P)\ :\ P\in\mathcal{P}_f(S,t_0) \,\}.
\end{align}
}

\section{Proposed global capacity and buffer management solution}
\label{sec: global management}

Our solution builds upon bundle-dependent \gls{sr}-\gls{cgr} \cite{SourceRoutBIRRANE2012} (see Section~\ref{ref_bundle_dep_sr}). We address buffer and capacity constraints during the initial route-search phase. This proactive approach contrasts with existing reactive solutions that address issues after they occur.

Our approach guarantees that all discovered routes satisfy buffer and capacity constraints. 
The modifications also preserve all viable solutions, ensuring the discovery of an optimal solution in the set of valid routes (\textcolor{blue}{see Section~\ref{sec_opti} for the proof}).

The system then encodes the optimal route, which respects all buffer and capacity limits, into the bundle for network forwarding, following the standard source-routing \gls{cgr} approach.

\subsection{Model-parameter modifications}
Our solution consists of two key elements:

\begin{itemize}
\item We modify the \gls{cp} through contact splitting for use in the route-search algorithm (i.e., Dijkstra search). The \gls{cp} modifications fall into two categories:
  \begin{itemize}
  \item Definitive modifications shared across the network.
  \item Temporary and bundle-specific modifications applied locally during the route search.
  \end{itemize}
\item Second, we restrict feasible edges between contacts through specific rules. The route-search algorithm enforces these rules during local route computation for each bundle.
\end{itemize}

\subsubsection{Contact splitting to modify the CP}
\label{sec: ct splitting intro}
We assume that the \gls{cp} is shared by all entities performing a route search (e.g., source nodes). 

\begin{figure}[!t]
	\centering
	\includegraphics[width = \columnwidth]{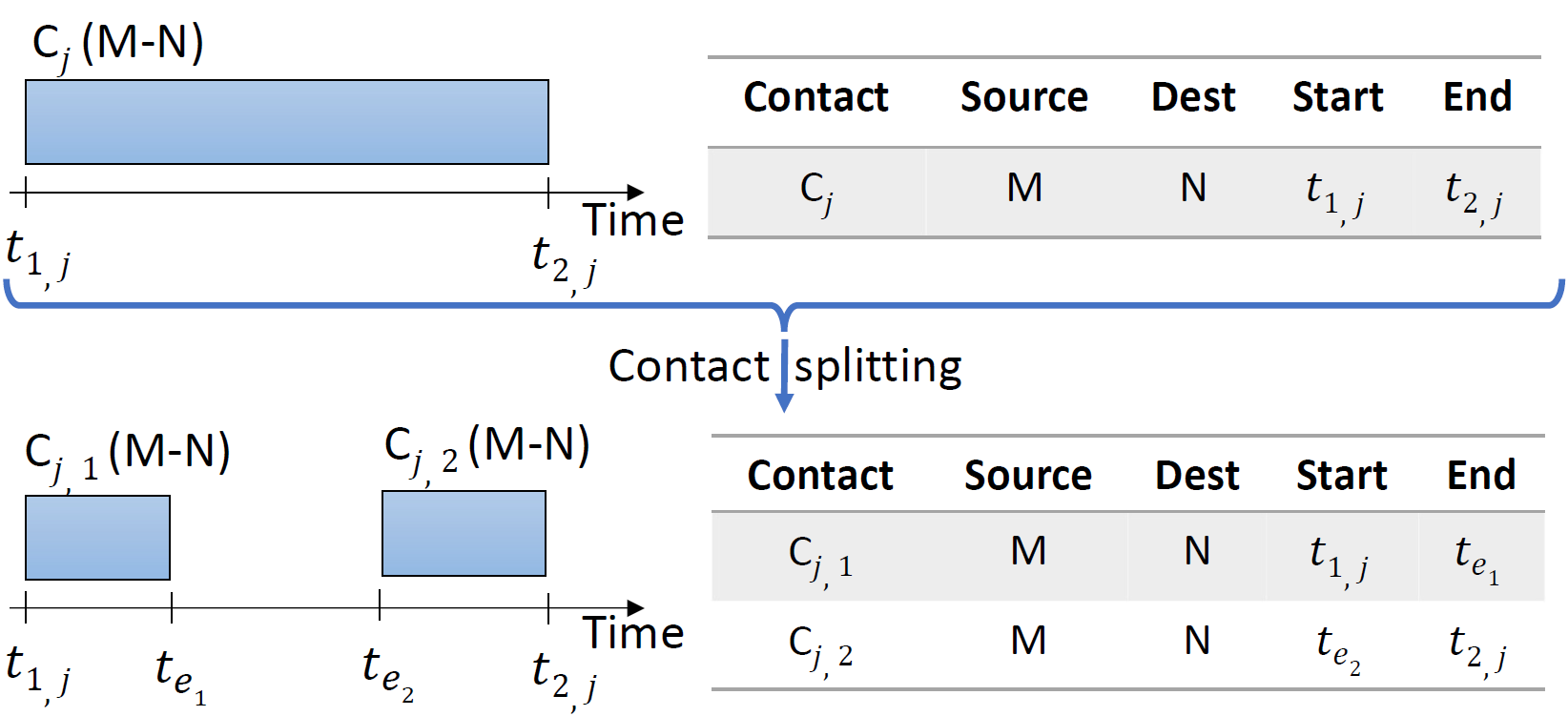}
	\caption{Contact splitting.}
	\label{fig: Contactsplit}
\end{figure}

We introduce the \emph{contact splitting} operation. Consider contact $\text{C}_j$ with data rate $R_j$ between nodes $M$ and $N$, active from $t_{1,j}$ to $t_{2,j}$. Contact splitting divides this contact into two new contacts $\text{C}_{j,1}$ and $\text{C}_{j,2}$ between $M$ and $N$: one from $t_{1,j}$ to $t_{e_1}$ and another from $t_{e_2}$ to $t_{2,j}$, where $t_{1,j} \leq t_{e_1} \leq t_{e_2}\leq t_{2,j}$ (see Figure \ref{fig: Contactsplit}). Both split contacts maintain the original data rate $R_j$.

When $t_{e_2} - t_{e_1} > 0$, the operation erases part of the original contact. Additionally, if either time span equals zero, contact splitting results in contact \emph{shortening}. 

\textcolor{blue}{As a result, contact splitting (for capacity and for buffer) turns $V_c^0$ into the vertex set $V_c$ of split contacts.}

\subsubsection{Edge pruning}
\label{sec_edge_prunning_prez}

\textcolor{blue}{We define \emph{edge pruning} as the act of removing from the connectivity-only graph $(V_c,E_c^0)$ the edges that would violate buffer constraints. The remaining edge set is $E_c\subseteq E_c^0$.
Edge pruning applies to any pair of contacts, whether original or split. For example, forbidding a given contact succession (e.g., the edge between $\text{C}_1$ and $\text{C}_2$) prevents their sequential use in any route.}

Edge pruning occurs during the route search, specifically within the \gls{crp}. The implementation adds a new criterion to the contact exploration process (lines $8-12$ in Algorithm \ref{alg:CRP}). When a contact fails this criterion, the algorithm:
\begin{itemize}
\item Skips it during exploration.
\item Prevents it from becoming a successor to the currently selected contact $\text{C}_s$.
\end{itemize}

In essence, edge pruning enforces route constraints by preventing invalid contact sequences during the search process.

\subsection{Capacity and buffer managements}

We explain how contact splitting (for capacity and for buffer) and edge pruning (for buffer) are used to address the capacity and buffer constraints discussed in Section \ref{sec:challenges}.

\textcolor{blue}{The main idea is that a new contact graph $(V_c,E_c)$ is built by contact splitting and edge pruning. It will be shown (Lemma 3, Section~\ref{sec_opti}) that the set of feasible routes $\mathcal{P}_f(S,t_0)$ coincides with the set of schedulable paths in $(V_c,E_c)$. Running CGR on $(V_c,E_c)$ then provides the optimal path $P^*$.}

\subsubsection{Capacity management}
\label{{sec:  proposedcapaManagement}}

\textcolor{blue}{In terms of the contact graph, capacity-feasibility enforcement uses contact splitting to update the vertex set: from the current set of contact segments $V_c^0$ to a new set $V_c$ in which the portions already consumed by routed bundles are removed.}

This approach requires knowledge of each bundle's size $S$. The algorithm begins by searching a route for bundle $i$. 
Similarly to the linear-volume approach, the algorithm excludes contacts with insufficient volume for the bundle.
Once the shortest path is selected for this bundle, the \gls{cp} is modified as follows:
We remove the part of the contact consumed by the bundle from the \gls{cp} before performing the route search for subsequent bundles. In the contact-graph view, \textcolor{blue}{this contact splitting updates $V_c^0$ to $V_c$ (only contact segments with sufficient remaining capacity remain in $V_c$). }
The length of the contact to be erased is $t_{\text{Tx,}i}^{\text{C}_j}$, starting from the contact start time $t_{1,j}$ or the arrival time of the bundle at the contact's sender node, whichever occurs later.
The transmission time $t_{\text{Tx,}i}^{\text{C}_j}$ for bundle $i$ over contact $\text{C}_j$ is:
\begin{equation}
	t_{\text{Tx,}i}^{\text{C}_j} = \frac{S_i}{R_j}.
\end{equation}

This contact splitting creates two new contacts (or one in case of shortening, see Section \ref{sec: ct splitting intro}) that replace the original one in the \gls{cp}. We apply this operation to all contacts in the selected route. All subsequent route searches use this modified \gls{cp} \textcolor{blue}{(i.e., the updated vertex set $V_c$)}. These modifications are permanent and must propagate to all nodes performing route searches.

\textbf{Example.} In Figure \ref{fig: volumeLinVSreal}, after the route search and the route selection for the initial bundle $a$ (striped rectangle), the \gls{cp} becomes as represented in Table \ref{tb: contact_plan_afterA}. The new \gls{cp} is then used for the route search of subsequent bundles. In this case, the part of contact $\text{C}_3$ between $t_i$
and $t_i+t_\text{Tx,a}^{\text{C}_3}$ is not an option for future bundle and no collision will therefore occur.  

\begin{table}
	\caption{Modified Contact Plan after allocating route for bundle $a$.}
	\label{tb: contact_plan_afterA}
\begin{center}
	
\begin{tabular}{|c | c |c |c |c|}
		\hline
		\textbf{Contact} &\textbf{Sender} & \textbf{Receiver} & \textbf{Start time}  & \textbf{End Time}  \\ \hline
		C$_{1,1}$ & A & B & $t_{1,1}$ &$t_{a}$ \\ \hline 
		C$_{1,2}$ & A & B & $t_{a} + t_{\text{Tx,}a}^{\text{C}_1} $ &$t_{2,1}$ \\ \hline  
		C$_{2,1}$ & B & C & $t_{1,2} + t_{\text{Tx,}a}^{\text{C}_2}$ &$t_{2,2}$  \\ \hline
		C$_{3,1}$ & C & D & $t_{1,3}$ &$t_{i}$   \\ \hline
	    C$_{3,2}$ & C & D & $t_{i} + t_{\text{Tx,}a}^{\text{C}_3} $ &$t_{2,3}$   \\ \hline
		C$_4$ & D & E  & $t_{1,4}$ &$t_{2,4}$   \\ \hline
	\end{tabular}

\end{center}
\end{table}

\textbf{Complexity.} The \gls{cgr} complexity is directly related to the size of the \gls{cp}. Contact splitting has two possible outcomes: either shortening (replacing with one contact) or splitting (replacing with two contacts). Each splitting operation increases the \gls{cp} size by at most one: In a full split, one original contact becomes two new contacts. Shortening preserves the \gls{cp} size by modifying the original contact's duration. The \gls{cp} size decreases by one when a shortened contact's volume falls below the minimum bundle size.
For bundle $i$, the number of split contacts equals the number of hops in its route $N_{\text{hops},i}$. In a time window $w$ with $N_{b,w}$ routed bundles, the maximum number of added contacts to the \gls{cp} is therefore:
\begin{equation}
\label{eq: deltaCPcapa+}
\Delta_{\text{CP},w}^+ = \sum_{i=1}^{N_{b,w}} {N_{\text{hops},i}}.
\end{equation}
Our implementation removes contacts (split or original) from the \gls{cp} upon expiration. Let $N_{\text{exp},w}$ represent the number of expired contacts in time window $w$. The resulting reduction in \gls{cp} size is simply $\Delta_{\text{CP},w}^- = N_{\text{exp},w}$.

Finally, the net change in the \gls{cp} size during window $w$ is:
\begin{equation}
\label{eq: deltaCPcapa}
\Delta_{\text{CP},w} =\Delta_{\text{CP},w}^+ - \Delta_{\text{CP},w}^-.
\end{equation}
Section \ref{sec: results} shows the algorithm's impact on the \gls{cp} size, and thus on the \gls{cgr} complexity, through practical examples.

\subsubsection{Buffer management}
\label{sec:  proposedBuffManagement}

We assume that each node has a maximum buffer capacity $B_\text{max}$, with overflow occurring beyond this limit. 
Our buffer management solution applies temporary and bundle-specific \gls{cp} modifications. Edge pruning follows the same temporary and bundle-specific pattern. These modifications remain local to the computing node and apply only during individual route searches.

\textbf{Forecast buffer tables.} 
We introduce forecast buffer tables as a key component of our solution.
Once a route is selected, the buffer tables are updated accordingly and shared to all entities performing a route search. \textcolor{blue}{These tables record, for each node $n$ and time $t$, the residual buffer capacity $\bar B_n(t)$, or equivalently the booked load $Q_n(t)$. Similarly to the \gls{cp}, all route-computing entities share these tables. They guide temporary modifications to both the \gls{cp} and allowed feasible edges.} 
In other words, once a route is validated, one checks when the bundle will reach a certain node and how long will it stay in its buffer 
according to the validated route. 

Since the buffer tables are updated as soon as a route is validated, all routing nodes know in advance the status of the buffers as function of time, even before the bundles travel through the network.

\textbf{Algorithm to respect the buffer constraints.} The proposed algorithm enforces buffer-limit constraints through the following approach:
Using bundle size and forecast buffer tables, the algorithm verifies buffer availability at each node throughout the route. The algorithm excludes routes that would cause buffer overflow at any node. First, it modifies the \gls{cp} temporarily through contact splitting \textcolor{blue}{(so $V_c^0$ becomes $V_c$)}. Next, it prunes specific feasible edges during the route search \textcolor{blue}{(removing from $E_c^0$ the edges that would cause overflow)}. \textcolor{blue}{This algorithm can be formulated with the four following main steps, see also Algorithm~\ref{alg:bufferSteps} for a summary. Then, the route search runs on $(V_c,E_c)$.}

\begin{algorithm}[t]
\color{blue}
\caption{\textcolor{blue}{Buffer-constraint enforcement (summary)}}
\label{alg:bufferSteps}
\begin{algorithmic}[1]
\Statex \textbf{Input:} bundle $b$ (size $S$), forecast buffer tables, \gls{cp} $(V_c^0,E_c^0)$
\Statex \textbf{Output:} feasible graph $(V_c,E_c)$ for route search
\State \textbf{Step 1:} Virtually add bundle size $S$ to all node buffer tables and detect potential overflows (nodes $\mathcal{Z}_\text{b}$, overflow sets $\mathcal{O}_{Z_k}$, times $t_{1,o_l}$, $t_{2,o_l}$).
\State \textbf{Step 2:} Temporarily modify \gls{cp}: remove (via contact splitting) contact portions that would forward $b$ to any $Z_k \in \mathcal{Z}_\text{b}$ during an overflow interval and obtain vertex set $V_c$.
\State \textbf{Step 3:} Identify problematic contact successions (first contact delivers bundle to node before overflow, successor contact starts after overflow begins).
\State \textbf{Step 4:} Prune from $E_c^0$ the edges corresponding to those successions. Route search runs on $(V_c,E_c)$ (enforced in \gls{crp}, Algorithm~\ref{alg:CRP}, lines 8--12).
\end{algorithmic}
\end{algorithm}

\begin{figure}[!t] 
		\centering
		\includegraphics[width=\columnwidth]{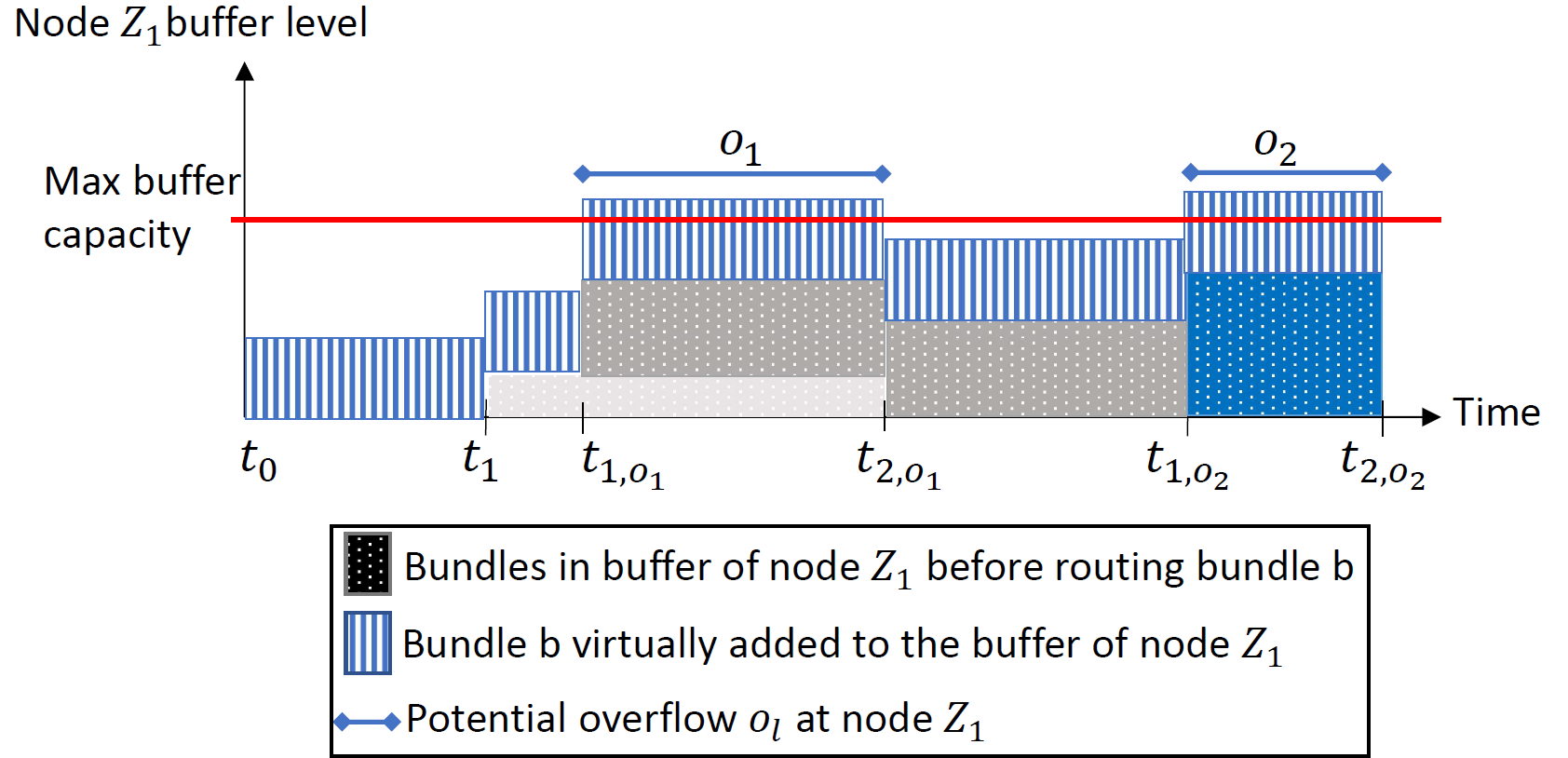}
		\caption{Potential overflows at node $Z_1$.}
		\label{fig: Z1 buff OF}
	\end{figure}

\begin{figure}[!t] 
		\centering
		\includegraphics[width=\columnwidth]{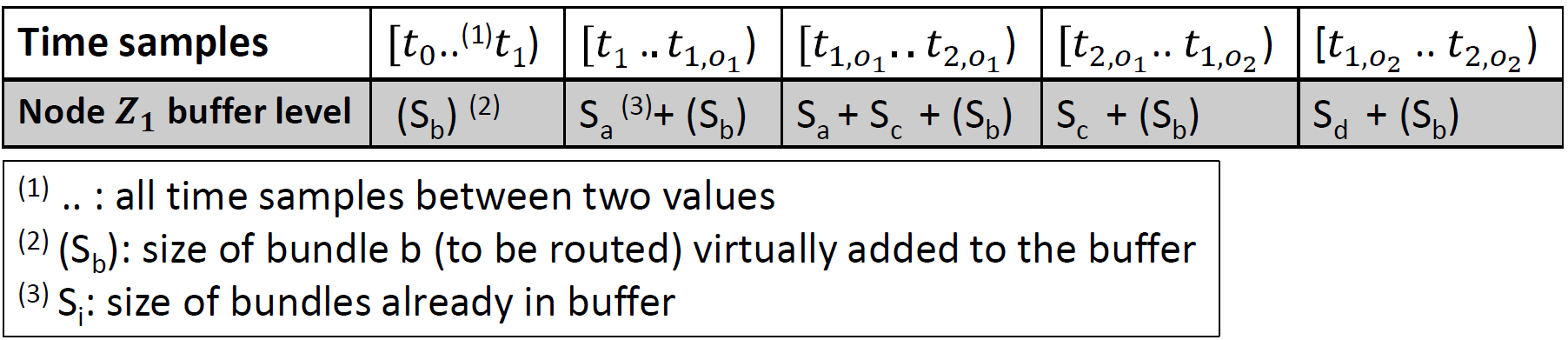}
		\caption{Forecast buffer table of node $Z_1$ before routing bundle b.}
		\label{fig: z forecast tb}
	\end{figure}

	\textbf{Step 1:} To begin with, the size of the bundle is virtually added to all node buffer tables at all times. 
	Potential overflows are detected and determined by their location (which nodes) and time (start and end times). 
	The set of nodes $Z_k$ with potential overflows for bundle b is denoted by $\mathcal{Z}_\text{b}$. 
	Each $Z_k \in \mathcal{Z}_\text{b}$ has a number of $n_\text{of} \geq 1$ potential overflows forming a set of overflows denoted by $\mathcal{O}_{Z_k}$. The start and end times of each overflow are denoted by $t_{1,o_l}$ and $t_{2,o_l}$, $o_l \in \mathcal{O}_{Z_k}$, respectively.  \\	
	\textbf{Example.} Figure \ref{fig: Z1 buff OF} presents an example where a bundle b arrives at a source node and needs to be routed. The figure depicts one node $Z_1 \in \mathcal{Z}_\text{b}$ having two potential overflows $o_1 \text{ and } o_2 \in \mathcal{O}_{Z_1}$. These potential overflows are identified by virtually adding the size of bundle b (vertically striped rectangles) to the buffer level of node $Z_1$ (dotted rectangles). The buffer level of node $Z_1$, before virtually adding the size of bundle b, is determined by validated routes of previous bundles (dotted rectangles). The corresponding forecast buffer table for node $Z_1$ is shown in Figure \ref{fig: z forecast tb}.

		\begin{figure}[!t] 
		\centering
		\includegraphics[scale = 0.38]{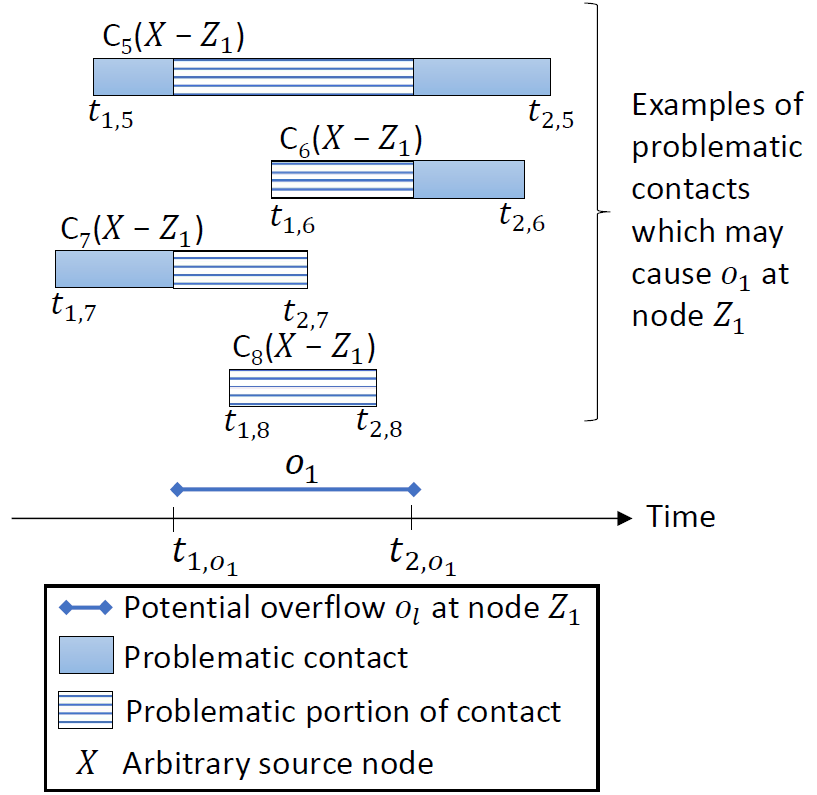}
		\caption{Contacts which might result in overflow $o_1$ at node $Z_1$.}
		\label{fig: pb contacts OF}
	\end{figure}
	
	\textbf{Step 2:} Then, the original \gls{cp} is temporarily modified accordingly (before starting the route search). One should remove all parts of the contacts from the \gls{cp} that might forward bundle b to any node $Z_k \in \mathcal{Z}_\text{b}$ between overflows start and end times using contact splitting. These problematic contacts are contacts that start before the end of an overflow and end after its start.\\ 
\textbf{Example.} Some problematic contacts for the potential overflow $o_1$ described in the previous example are shown in Figure \ref{fig: pb contacts OF}. These contacts have node $Z_1$ as receiver (regardless of their sender node $X$), start before $t_{2,o_1}$, and end after $t_{1,o_1}$. They are problematic because their duration intersects the period of overflow $o_1$. The problematic portions of these contacts are horizontally striped in Figure \ref{fig: pb contacts OF}. They may forward bundle b to $Z_1$ during the potential overflow $o_1$ duration. The parts of these contacts that intersect the overflow duration are to be temporarily removed using contact splitting. The same applies to all potential overflows at all nodes.  

\textbf{Step 3:} Removing the problematic contact's potential overflow parts is not sufficient. One should also address problematic contact successions. These problematic contact successions occur between contacts having an end time before the beginning of the potential overflow $o_i$ and a succeeding contact in a route having a start time after the beginning of overflow $o_i$. Such succession of contacts is problematic because it results in having the bundle in the buffer of the node during the potential overflow. This step therefore consists in identifying the problematic contact successions as described via the following example.\\
\textbf{Example.} After step 2, there is no contact having $Z_1$ as receiver that spans the duration of $o_1$ in Figure \ref{fig: pb contacts OF}. Nevertheless, there are contacts having $Z_1$ as receiver and ending before $t_{1,o_1}$. Using one of these contacts (e.g., $\text{C}_i$), a bundle arrives to $Z_1$ before the potential overflow $o_1$. However, this bundle must leave $Z_1$ before the beginning of overflow $o_1$ to avoid the potential problem. 
The bundle leaves node $Z_1$ through a contact succeeding the previous one ($\text{C}_i$) in a route. The succeeding contact (e.g., $\text{C}_j$) has $Z_1$ as sender, regardless of the receiver. 
As a result, the succeeding contact $\text{C}_j$ should start before the potential overflow $o_1$ to avoid it. Otherwise, the succession of $\text{C}_i$ and $\text{C}_j$ in a route causes the overflow and is therefore problematic.

	\textbf{Step 4:} Finally, the last step is to remove from $E_c^0$ the edges between contacts that yield problematic contact successions, so that the route search runs on the feasible graph $(V_c,E_c)$.
In other words, feasible edges for which the succeeding contact starts too late are pruned. 
This edge pruning is performed during the route search, in the \gls{crp}: lines $8-12$ in Algorithm \ref{alg:CRP}. Note that additional inputs were required for the \gls{crp} to make our edge pruning possible: line $2$ in Algorithm \ref{alg:CRP}. In addition to its original inputs, the \gls{crp} takes the bundle $b$ and the overflows start times$T_{1,s}$ at the receiver of the selected contact $\text{C}_s$, respectively. \\
	\textbf{Example.} In the example of Figure \ref{fig: pb contacts OF}, during the route search, if a contact $\text{C}_s$ forwards the bundle to node $Z_1$ before $o_1$, the edge between $\text{C}_s$ and a contact $\text{C}_i$ from $Z_1$ to a third node is only accepted if $\text{C}_i$ starts before $ t_{1,o_1}$. The same applies to all overflows at all nodes. 
More precisely, let us consider, in a \gls{crp} iteration, that $\text{C}_s$ is the selected contact and $\text{C}_i$ is another contact in the \gls{cp}.  
Also, let us consider that the receiver of $\text{C}_s$ is node $Z_1$, having one potential overflow $o_1$. To consider $\text{C}_i$ as explored or a candidate successor for $\text{C}_s$, it must pass all the checks in the original \gls{crp} (lines $5-7$ in Algorithm \ref{alg:CRP}). In addition to these checks, we add the following condition: If $f_s$, the bundle arrival time to the receiver of $\text{C}_s$, is before $ t_{1,o_1}$ ($f_s < t_{1,o_1} $), then the start time of $\text{C}_i$ must also be before $ t_{1,o_1}$ ($t_{1,i} < t_{1,o_1} $). Otherwise, $\text{C}_i$ is ignored in the \gls{crp} and it is not explored. If this condition is not met, $\text{C}_i$ cannot be a candidate successor for $\text{C}_s$. In a more general scenario, where multiple potential overflow periods exist at a node buffer, a bundle arriving at the node via contact $\text{C}_s$ between two overflow periods should depart from the node using contact $\text{C}_i$ before the next overflow period begins (lines $8-12$ in Algorithm \ref{alg:CRP}).

As described by the above steps, only invalid routes are removed and no feasible solution is omitted. \textcolor{blue}{Optimality of the route returned by \gls{cgr} is formalized in Section~\ref{sec_opti}.}

\textbf{Complexity.} 
As already mentioned, the \gls{cgr} complexity directly depends on the \gls{cp} size. 
Buffer management temporarily increases the \gls{cp} size when routing bundle $b$, based on the bundle's size and current buffer occupancy. The size increase depends on two factors: the number of overflowed nodes $|\mathcal{Z}_\text{b}|$ and the contacts intercepting overflows at these nodes. Each node $Z_k$ can experience multiple overflows, denoted as $|\mathcal{O}_{Z_k}|$. When a contact intersects an overflow period, it splits into at most two new contacts, increasing the \gls{cp} size by one. This increase remains temporary and applies only during bundle $b$'s routing phase. After routing completes, the system reverts to the permanent \gls{cp}.
At most, the number of temporarily added contacts to the \gls{cp} for routing bundle $b$ with buffer constraints is:
\begin{equation}
\Delta_{\text{CP},b}^+ = \sum_{k=1}^{|\mathcal{Z}_\text{b}|} \sum_{l=1}^{|\mathcal{O}_{Z_k}|} {N_{\text{c}_{l,k}}},
\end{equation}
where $N_{\text{c}_{l,k}}$ represents the number of contacts having node $Z_k \in \mathcal{Z}_\text{b}$ as destination and intercepting overflow $o_l \in \mathcal{O}_{Z_k}$. Section \ref{sec: results} shows the practical impact of this management solution on the \gls{cp} size.

\textcolor{blue}{\subsection{Optimality of the proposed approach}\label{sec_opti}}



\medskip


\medskip

\textcolor{blue}{\noindent\textbf{Lemma 1 (capacity enforcement).}
The capacity enforcement (contact splitting for capacity) modifies the contact graph so that a route is excluded (i.e., is no longer a path in the graph, because it would use a contact portion removed by splitting) iff it is capacity-infeasible (i.e., does not respect the contact-capacity constraint defining $\mathcal{P}_f$): (i) every route excluded in this way would violate that constraint, and (ii) every route that would violate it is excluded.}

\textcolor{blue}{\emph{Proof.}
(i) If a route is excluded by the capacity enforcement, it would use at least one contact portion that was removed by contact splitting for capacity (already allocated to prior bundles). By construction, any such route would exceed available contact capacity. So every route excluded in this way is capacity-infeasible (does not respect the contact-capacity constraint defining $\mathcal{P}_f$).
(ii) If a route would violate the contact-capacity constraint, then it would use a portion of some contact that is already allocated to another bundle. That contact portion is removed by contact splitting for capacity. Thus every capacity-infeasible route is excluded (no longer a path in the graph). $\square$}

\medskip

\textcolor{blue}{\noindent\textbf{Lemma 2 (buffer enforcement, Algorithm~\ref{alg:bufferSteps}).}
The buffer enforcement builds the feasible graph $(V_c,E_c)$: contact splitting for buffer updates the vertex set $V_c$, and edge pruning yields $E_c\subseteq E_c^0$. A route is excluded (i.e., is not a path in $(V_c,E_c)$) iff it is buffer-infeasible (i.e., does not respect the node-buffer constraint defining $\mathcal{P}_f$): (i) every route excluded in this way (it uses a contact portion removed by buffer splitting or an edge pruned for buffer) would violate that constraint, and (ii) every route that would violate it is excluded.}

\textcolor{blue}{\emph{Proof.}
(i) The buffer procedure builds $(V_c,E_c)$ via contact splitting (Step~2) and edge pruning (Step~4). A route is excluded (not a path in $(V_c,E_c)$) only if it would use a contact portion forwarding $b$ to some $Z_k\in\mathcal{Z}_\text{b}$ during an overflow interval, or an edge corresponding to a problematic succession. By construction, any such route would exceed $\bar B$ at some node and time. So every route excluded in this way is buffer-infeasible (does not respect the node-buffer constraint defining $\mathcal{P}_f$).
(ii) If a route would violate the node-buffer constraint at some node $Z$, then $Z\in\mathcal{Z}_\text{b}$ and the route either forwards $b$ to $Z$ during an overflow interval (hence that contact portion is removed in Step~2) or uses a problematic succession at $Z$ (hence the corresponding edge is pruned in Step~4). Thus every buffer-infeasible route is excluded (not a path in $(V_c,E_c)$). $\square$}

\medskip

\textcolor{blue}{\noindent\textbf{Lemma 3 (bijection).}
A route is a \emph{feasible route} (i.e., $P\in\mathcal{P}_f(S,t_0)$ as defined in Section~\ref{sec_math_sub})
iff it is a schedulable vertex path in $(V_c,E_c)$.}

\textcolor{blue}{\emph{Proof.}
($\Rightarrow$) A feasible route is, by definition, in $\mathcal{P}_f(S,t_0)$, hence capacity-feasible and buffer-feasible. By Lemmas 1 and 2, a route is excluded (no longer a path in the graph / not a path in $(V_c,E_c)$) iff it violates the contact-capacity or node-buffer constraint defining $\mathcal{P}_f$. So a feasible route is not excluded and uses only contact portions and successions that remain in $(V_c,E_c)$. Its vertices therefore lie in $V_c$ and its edges in $E_c$, and with its per-contact start times it is schedulable. Thus the route induces a schedulable vertex path in $(V_c,E_c)$.
($\Leftarrow$) A schedulable vertex path in $(V_c,E_c)$ uses only vertices in $V_c$ and edges in $E_c$, so it was not excluded by the capacity or buffer enforcement. By Lemmas 1 and 2, it is therefore capacity-feasible and buffer-feasible. With schedulability it lies in $\mathcal{P}_f(S,t_0)$, i.e., it is a feasible route. $\square$}

\medskip

\textcolor{blue}{In the framework of \cite{Dean2004TDSP}, the following holds.}

\medskip

\textcolor{blue}{\noindent\textbf{Lemma 4 (FIFO time-dependent shortest path).}
On a FIFO, nonnegative time-dependent graph with waiting allowed,
time-dependent Dijkstra returns a path of minimum arrival time
for a fixed departure time.}

\medskip

\textcolor{blue}{\noindent\textbf{Theorem (optimality).}
Let $P^*$ be the vertex path returned by time-dependent Dijkstra on $(V_c,E_c)$.
Then $A(P^*)=\min\{\,A(P): P \in \mathcal{P}_f(S,t_0)\,\}$.}

\textcolor{blue}{\emph{Proof.}
By Lemma 4, $P^*$ minimizes arrival time among all vertex paths in $(V_c,E_c)$.
By Lemma 3, those vertex paths are exactly the feasible routes (elements of $\mathcal{P}_f(S,t_0)$). Hence $P^*$ is optimal. $\square$}


\subsection{Booking resources for forwarding safety margin}
\label{sec: safety margin}

In the proposed solution, only source nodes monitor buffer and contact capacity status. This centralized monitoring solution enables source nodes to prevent overflows and overbooking. However, unpredictable disruptions (such as antenna pointing errors or node power outages \cite{UCPMadoery2018}) may force intermediate nodes to perform routing during the forwarding phase. Intermediate nodes lack access to current buffer and capacity status, risking overflows or overbooking with their routing decisions. To address this risk, we reserve portions of contact and buffer resources, keeping them unused by source nodes. These reserved portions, called safety margins, allow intermediate nodes to reroute bundles safely without causing overflows or overbooking.

The network's disruption level determines the required resource allocation. This allocation requires careful tuning to optimize resource usage. The safety margin must balance two requirements: sufficient size to prevent bundle loss during intermediate node rerouting, while remaining minimal to preserve network efficiency.

These safety margins can be implemented as follows. 
First, regarding the contact-capacity management, two \glspl{cp} are created.
Contact splitting is performed for the slots that should be kept as margin.
The safety margin is therefore removed from the original \gls{cp}.
The obtained \gls{cp} is provided to and used by the source nodes. 
The safety-margin part (contacts removed from the original \gls{cp}) is forwarded to intermediate nodes that may perform routing in unpredictable situations. 
In terms of buffer management, source nodes utilize a fraction of the buffer capacity, leaving the rest available for intermediate nodes. This solution therefore enhances robustness to potential unexpected events, despite the poor knowledge of the current network status by intermediate nodes.

\subsection{Information sharing in the network}

The modified network information after contact splitting (except the bundle-specific \gls{cp} modifications) and the modified forecast buffer tables (after allocating a route for a bundle) are to be shared with all source nodes performing a route search. This is similar to the sharing of the up-to-date \gls{cp} with the source nodes in standard SR-\gls{cgr}. 
The safety margin resources are shared with all nodes in the network at a low frequency.

\section{Benchmark: source-routing CGR}
\label{sec: benchmark}

\textcolor{blue}{To our knowledge, no existing benchmark was available under the centralized, perfect-CP assumptions we consider. We therefore describe our source-routing CGR benchmark in this section.}

As mentioned in the first part of the document, network issues are handled after the route search in standard \gls{sr}-\gls{cgr}.
While re-transmission mechanisms do exist for long round-trip times communication, such as \gls{ltp} \cite{LTP2015}, the re-transmission approach in case of failure in standard \gls{sr}-\gls{cgr} is not well described in the literature.
As there are many manners to address these issues, we describe in details the assumptions made for the considered \gls{sr}-\gls{cgr} benchmark. The proposed algorithm is compared to this benchmark in Section \ref{sec: results}, where the simulation results are presented.

To begin with, both in the benchmark and in our algorithm, when a bundle is being transmitted over a contact, it is assumed that the entire link is occupied for the time required to transmit the bundle. In other words, the whole link capacity is used (i.e., \gls{tdm} approach). 
The initial \gls{cp} is constructed at the source and is broadcasted to all nodes in the network.

Then, the source node performs a route search using Yen's algorithm with Dijkstra searches. A list of the $k$ (Yen's constant) shortest routes is generated. This list is kept until a change occurs in the \gls{cp}, such as the end of a contact. In this case, the route list is discarded and a new list is generated, as in route-list pruning \cite{SABRspecCCSDS2019}. A list of candidate routes is constructed as explained in Section \ref{sec: valid}. When no candidate route is found in this list, the bundle is delayed and the route search is performed again after one time step. 

Regarding the capacity and buffer managements performed before the route search: When a route is found, the volumes occupied by the bundle in the selected route are deducted from the available volume of corresponding contacts in the \gls{cp} (i.e., linear volume approach as described in Section \ref{sec: capa challenges}). 
No buffer information is available or considered at the source.

Finally, the route is encoded in the bundle and the bundle is sent in the network. 
A bundle with an encapsulated route at a transmitting node (source or intermediate node) is subject to checks to verify whether this route can be used or not. 
The contact capacity and buffer managements during the forwarding phase in this benchmark are detailed hereafter.

\subsubsection{Contact-capacity management after the route search} 

First, we verify the time at which the corresponding contact must be used for the bundle transmission. In the case of collision with another bundle, the transmission of the bundle having the highest ID is delayed (similar behavior to that of determining the \gls{eto}). The ID increases with the bundle generation time at the source. Two cases can occur:
\begin{itemize}
\item If delaying the bundle does not cause any problem, the times of future hop uses are simply updated in the encapsulated route according to the delay caused by other bundles using the same contact. This update accounts for the delay caused by other bundles using the same contact. 
\item If delaying the bundle causes its transmission at a time later than the corresponding contact's end time, a new route is computed from the node where the problem happened, avoiding this problematic contact. 
The time of the new route-search call is the problem-detection time.
\end{itemize}
We provide further details regarding the second case: 
To reroute the bundle, the up-to-date \gls{cp} at the problem detection time is loaded instantly from the source node. Note that this is an advantageous assumption for the benchmark. Moreover, since the remaining part of the old route is no longer used, the volumes of future contacts in the old route are updated locally in the \gls{cp} at this node (i.e., incrementing the available volume of the unused contacts by the size of the bundle). \gls{cp} updates performed on intermediate nodes are not reported to the source. The newly computed route, if any, is encapsulated within the bundle for further forwarding. 

If none of the $k$ shortest routes is valid, the bundle is sent instantly to the source and rerouting from the source is performed after one time step. 

 \subsubsection{Buffer management after the route search} During forwarding phase, we assume that a node only has access to the state of its buffer and that of its neighbors (nodes in contact). 
When a bundle is to be transmitted, the current node verifies if there is enough room in the buffer of the receiving node. If this is the case, the next node's buffer occupancy is increased to book space for the bundle and the bundle is transmitted. 

A waiting time is evaluated to determine how long the bundle will remain in the buffer of the next node. 
This time is equal to the transmission time from the current node to the next node $t_\text{tx,1}$, as well as the wait time until the start of the next contact (if this wait exists) $t_\text{wait}$ and the transmission time from the next node to a more distant node in the route $t_\text{tx,2}$. During $t_\text{tx,1} + t_\text{wait}$, the total bundle size is assumed to occupy the buffer of the next node. 
During $t_\text{tx,2}$, the space occupied by the bundle is linearly reduced as time passes (i.e., as the bundle leaves the buffer) according to the data rate of the contact used for this transmission. 

If there is no room for the bundle in the buffer of the next node, the bundle is rerouted from the current node via a route that avoids this problematic neighbor as well as previously visited nodes to avoid loops. Note that a node whose buffer is fully occupied is considered problematic and must therefore be avoided. 
To do this, the problematic contact that was to be used in the route is removed from the \gls{cp} for the rerouting of this bundle from this node only. Note that the bundle can pass through the problematic node later in the route, if the buffer has been emptied for example. 

If no valid route is found among the $k$ shortest routes, the bundle is instantly sent back to the source and rerouted after one time step. 

An alternative solution, which we are not considering, would be to continue trying to transmit via the fully booked node at each time step until a valid route is found or a place becomes available in the buffer of the problematic node, whichever comes first. 
Note that this solution requires a lot of computing resources as the route search is repeated at each time step until a suitable solution is found.

\section{Simulation environment and setup}

\label{sec: setup}

\subsection{Environment}

In the simulations, we consider $N_\text{sat}$ satellites orbiting in $N_\text{p} =4$ orbital planes distributed according to a Walker-delta constellation \cite{Walker1984} with zero phasing. The orbital planes are evenly distributed in the range $[0\degree \ 360\degree]$. Hence, there is an angle $ \Delta \Omega = \frac{360\degree}{N_\text{p}}$ between the \gls{raan} of two satellites in adjacent planes. Furthermore, on each plane, the true anomalies of $\frac{N_\text{sat}}{N_\text{p}}$ satellites are evenly separated by $\frac{360}{N_\text{sat}/N_\text{p}} \degree$ . We consider circular orbits ($\text{eccentricity} = 0$) with an inclination of $52\degree$ and a radius of $(6371 + 780) \times 10^3$ m, where $6371 \times 10^3$ m is the Earth radius constant. 

We consider two \glspl{gs} (\gls{gs}1 and \gls{gs}2). Bundles are generated by \gls{gs}1, which is the source. The bundles are to be transmitted from this source to \gls{gs}2, the destination. The satellites are the intermediate nodes of the network.

The \gls{cp} is constructed for a 24-hour scenario starting from 8:55 pm on August 19, 2020. We assume that all the contacts have the same rate $R_j$.

\begin{figure}[!t] 
	\centering
	\includegraphics[trim = 1.6in 1.8in 4.8in 1.5in, clip, width=\columnwidth]{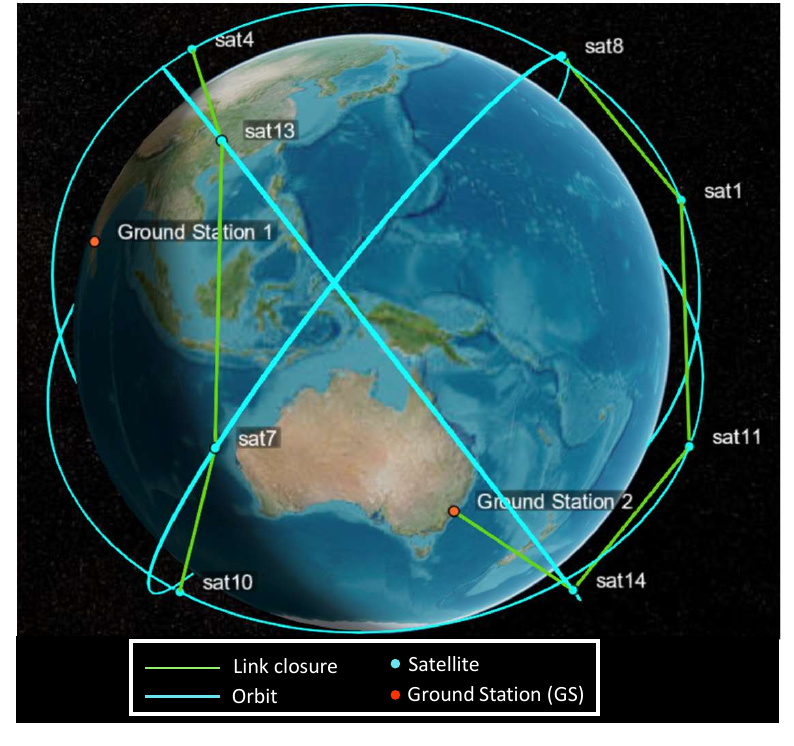}
	\caption{Snapshot of the satellite communication scenario with $N_\text{sat}=16$ at 11:18:08 pm on August 19, 2020. This corresponds to constellation 3 in Table~\ref{tb: ConstChar}.}
	\label{fig: sat scenario}
\end{figure}

A snapshot of the \glspl{gs} and one Walker constellation with $N_\text{sat}=16$ at 11:18:08 pm on August 19, 2020 is illustrated in Figure \ref{fig: sat scenario}. It was generated using MATLAB Satellite Communications Toolbox \cite{MatlabSatComTB}. 
Using cislunar constellations to evaluate the algorithm will be considered in future work (see also the conclusion).

In the considered constellations, each orbital plane has $\frac{N_\text{Sat}}{N_\text{p}}$~satellites. We define the plane hops as the number of planes a communication between two satellites can cross (Table \ref{tb: ConstChar}). Depending on the communication parameters and the satellite dynamics, a satellite located in an orbital plane can communicate with satellites located in the same plane (zero plane hop, same \gls{raan}), in adjacent planes (one plane hop, one $\Delta \Omega$ difference between \glspl{raan} of the satellites) or in more distant planes (more than one plane hops, meaning that there is no plane restriction in our case as the number of planes is limited).
	


The characteristics of the used constellations, also depending on the communication parameters provided in the following subsection, are given in Table \ref{tb: ConstChar}.

\subsection{Communication parameters}

Each satellite is mounted with transmit and a receive Gaussian antennas having a dish diameter of $0.5$ and an aperture efficiency of $0.65$. The transmission frequency and power are $30\times10^9$ Hz and $15$ dB, respectively. The receiver's gain to noise ratio and energy to noise sensitivity are equal to $3$ dB/Kelvin and $4$ dB, respectively. The source \gls{gs} is equipped with a transmit antenna whose properties are similar to those of satellites, except that the transmission power is considered to be $30$ dB. The destination \gls{gs} is equipped with a receive antenna whose properties are similar to those of the satellites, with the exception of its sensitivity of $1$ dB. A free space channel model is considered. The considered minimum required elevation angles of the \glspl{gs} for communication is $0 \degree$.

Two satellites can communicate with each other if they are in line-of-sight and if the power received by the receiving satellite from the transmitting satellite is greater than the sensitivity of the receiving satellite. For a satellite to communicate with a \gls{gs}, the elevation angle of the former relative to the latter must be greater than the minimum elevation angle of the latter, in addition to being in line-of-sight and within the communication range. If these conditions are met, there is link closure between two nodes. A contact extends from the beginning to the end of a link closure between two nodes. \\
\textbf{Example.} In Figure \ref{fig: sat scenario}, constellation $3$ is represented. The figure illustrates, in green, link closures between pairs of nodes: sat13-sat7, sat11-sat1, and sat14-\gls{gs}2, for example. These link closures happen at the time of the snapshot. The link between sat14 and \gls{gs}2 is at its limit: One time sample (of one second) later, this link no longer exists as these two nodes cease to respect the elevation angle constraint. The link closure is thus interrupted even though they remain in line-of-sight and in their communication range. In addition, sat4 and sat8 may be within communication range, but their communication is blocked by the Earth, preventing them from being in line-of-sight.

\begin{table}[h]
	\caption{Constellation characteristics.} 
	\label{tb: ConstChar}
\begin{center}
\resizebox{\linewidth}{!}{	
\begin{tabular}{|c|c|c|c|c|c|}
		\hline
		Constellation &\multirow{2}{*}{$N_\text{p}$} & \multirow{2}{*}{$N_\text{sat}$} & Initial number of  & Mean contact & Plane  \\
		number	&  & & contacts in \gls{cp} &  duration (s) & hops\\ \hline
		1 & $4$ & $4$ & $284$ & $518.91$ &  $1^\text{(a)}$ \\ \hline  
		2 & $4$ & $8$ &$802$ & $701.91$ & $2^\text{(b)}$  \\ \hline
		3 & $4$ & $16$ & $3478$ & $834.75$ & $2$  \\ \hline
		\multicolumn{6}{l}{$^\text{(a)}$ 1: A satellite can only communicate with satellites in same or  } \\
				\multicolumn{6}{l}{ \hspace{6mm} adjacent orbital planes to its own plane (up to one plane hop).} \\
		\multicolumn{6}{l}{$^\text{(b)}$ 2: A satellite can also communicate with satellites in orbital planes } \\
	  \multicolumn{6}{l}{ \hspace{6mm} adjacent to those adjacent to its own plane (up to two plane hops).}

	\end{tabular}}

\end{center}
\end{table}

\subsection{Other simulation parameters and assumptions}
\label{sec_sim_param}

At each time step, one bundle of $100$ Bytes is generated at the source. The bundles are generated over a period of $2000$ seconds, starting from the beginning of the scenario time, with equal inter-arrival times. 
In total $N_\text{b}$ bundles are generated and forwarded. 
This means that one bundle is generated each $\frac{2000}{N_\text{b}}$ seconds (the bundle inter-arrival time decreases as $N_\text{b}$, the total number of bundles generated at source, increases).



Yen's constant $k$ is chosen equal to $10$.
Finally, we assume that there is no external disruption in the network. That is, the provided \gls{cp} is valid and no rerouting is needed on intermediate nodes due to external factors. Given these assumptions, there is therefore no need to consider any safety margin for contact capacity and buffer space (Section \ref{sec: safety margin}).

\subsection{Simulation tools}

We use MATLAB Satellite Communications Toolbox \cite{MatlabSatComTB} to simulate the presented satellite communication scenario. We can then calculate the link closures between each pair of nodes as a function of time to obtain the \gls{cp} used in the benchmark and in our solution, also implemented in MATLAB.

An alternative algorithmic implementation would be to use \gls{ion} \cite{IONBurleigh2007} \cite{IONsoft}. However, we have not chosen it because our main focus is on \gls{cgr}.

\section{Simulation results}
\label{sec: results}

We evaluate the performance using two main metrics. The first metric is the average time a bundle spends in the network. This is computed as the difference between the destination-arrival time and the bundle-generation time at the source. It includes buffer waiting time, transmission time, and potential rerouting delays. The other metric is the number of rerouting events. For \gls{sr}-\gls{cgr}, it includes rerouting at the source when no route is found among the $k$ shortest ones as well as when a bundle is instantly sent back to this source. It also includes, both for \gls{sr}-\gls{cgr} and the proposed solution, rerouting at an intermediate node if a buffer or a capacity check fails.

\textcolor{blue}{In our idealistic simulation setting, the delivery ratio is one: all bundles eventually reach the destination as the simulation runs until all bundles are delivered (possibly after reroutes, as described in Section~\ref{sec: benchmark} for the benchmark). 
The impact of buffer size on performance (and thus on buffer utilization) is illustrated in the Figure~\ref{fig: timeInNet_Bvar} where buffer size is varied.}

We evaluate the impact of contact-capacity management and buffer management separately. Of course, the two management techniques can be used together to obtain greater gain. 




\subsection{Impact of contact-capacity management}

In this section, we only evaluate the performance of the proposed contact-capacity management. 
Therefore, we do not consider the constraints associated with buffer limits, i.e., the maximum node buffer capacity $B_\text{max} =~\infty$ for all nodes.

\begin{figure}[!t]
	\centering
	\includegraphics[trim = 1.1in 3.5in 1.5in 3.5in, clip, width=\columnwidth]{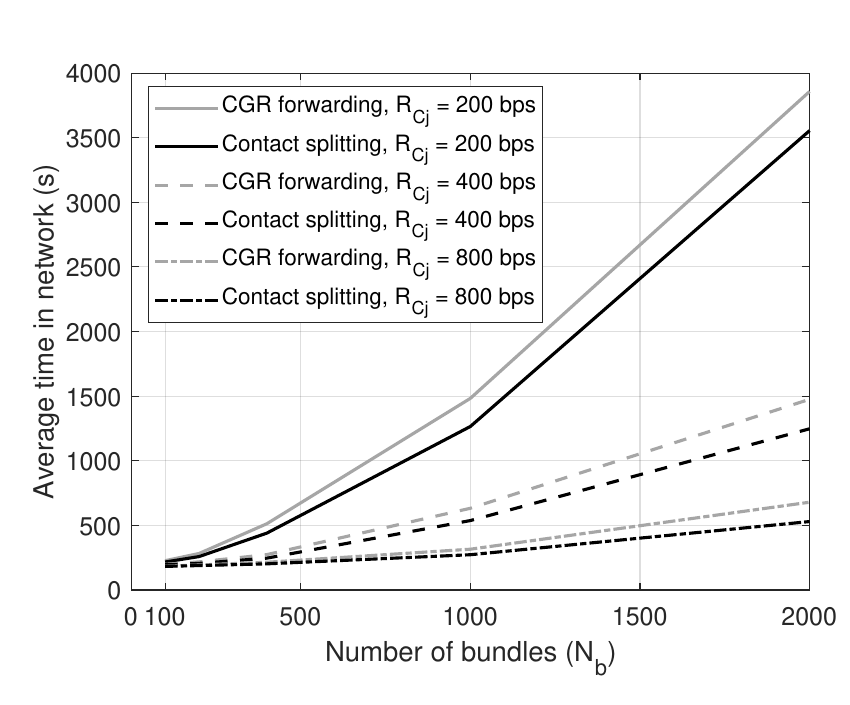}
	\caption{Average time spent in the network ($N_\text{sat} = 16$).}
	\label{fig: timeInNet_const1}
\end{figure}

\begin{figure}[!t]
	\centering
	\includegraphics[trim = 1.1in 3.5in 1.5in 3.5in, clip, width=\columnwidth]{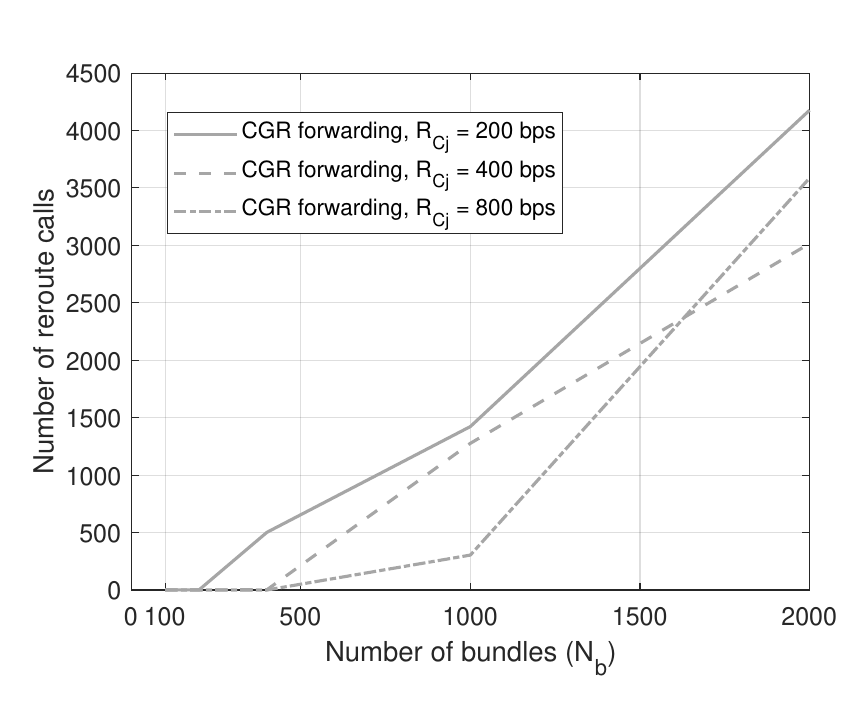}
	\caption{Number of reroute calls for CGR ($N_\text{sat} = 16$).}
	\label{fig: numrerouteConst1}
\end{figure}

Figure \ref{fig: timeInNet_const1} shows the average time spent in the network for $N_\text{sat} = 16$ (constellation $3$) as function of the number of bundles generated at source $N_\text{b}$. Three values of contact data rate $R_j$ are considered: $200$, $400$, and $800$ \gls{bps} for all contacts. As expected, one can see that the time spent in the network increases with the number of bundles and with decreasing data rate. With low load (e.g., $N_\text{b}<400$ for $400$ \gls{bps}), the proposed algorithm and the \gls{sr}-\gls{cgr} have similar performance. As $N_\text{b}$ increases, the gap between the curves increases, with less time spent in the network when using our proposed method. This is due to the increased number of rerouting events during forwarding time when using \gls{sr}-\gls{cgr}. This number of rerouting events is shown in Figure \ref{fig: numrerouteConst1} for the same $N_\text{sat}$ (constellation $3$) and $R_j$ values as function of $N_\text{b}$. With low load, there are only few overbooking events happening on the contacts. In this case, traditional \gls{sr}-\gls{cgr} has satisfactory performance. When the number of bundles increases, the overbooking phenomenon increases, making rerouting more likely. In case of such overbooking events, intermediate nodes try to overcome this issue by rerouting the bundle from their location toward the destination. Such new routes are longer than the ones avoiding overbooking, computed using the proposed algorithm. Moreover, if no route is found from the node, the bundle has to be rerouted from the source (see the assumptions on the benchmark) also causing additional delay.
The proposed algorithm is more efficient because it computes the shortest route directly from the source node while avoiding overbooking. 
Indeed, as expected, our proposed solution required no rerouting due to contact capacity check failures during the simulation.

Moreover, although in some cases the time spent in the network with \gls{sr}-\gls{cgr} is similar to the proposed solution, the intermediate node complexity is significantly higher with the benchmark due to multiple checks and rerouting events. 

\begin{figure}[!t]
    \begin{subfigure}[b]{0.5\textwidth}
     \includegraphics[trim = 1.1in 3.7in 1.5in 3.7in, clip, width=\columnwidth]{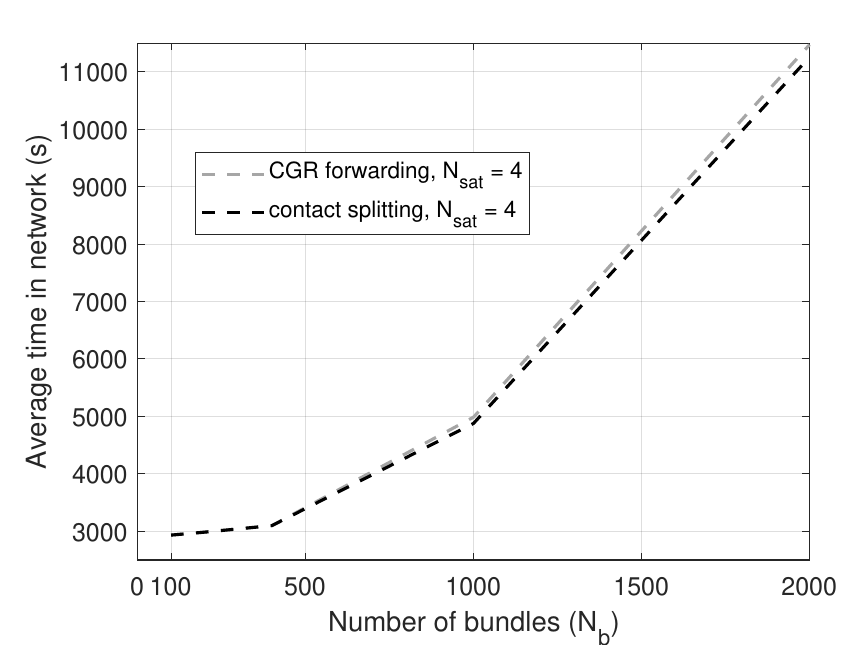}
	\caption{$N_\text{sat} = 4$.}
	\label{fig: timeInNet_4sats}
    \end{subfigure}
    \begin{subfigure}[b]{0.5\textwidth}
      \includegraphics[trim = 1.1in 4.3in 1.5in 4.5in, clip, width=\columnwidth]{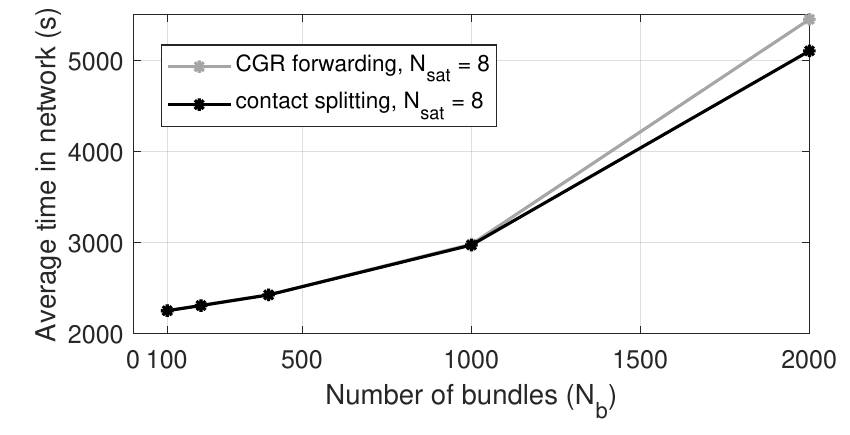}
	\caption{$N_\text{sat} = 8$.}
	\label{fig: timeInNet_8sats}
    \end{subfigure}
    \begin{subfigure}[b]{0.5\textwidth}
      \includegraphics[trim = 1.1in 4.3in 1.5in 4.5in, clip, width=\columnwidth]{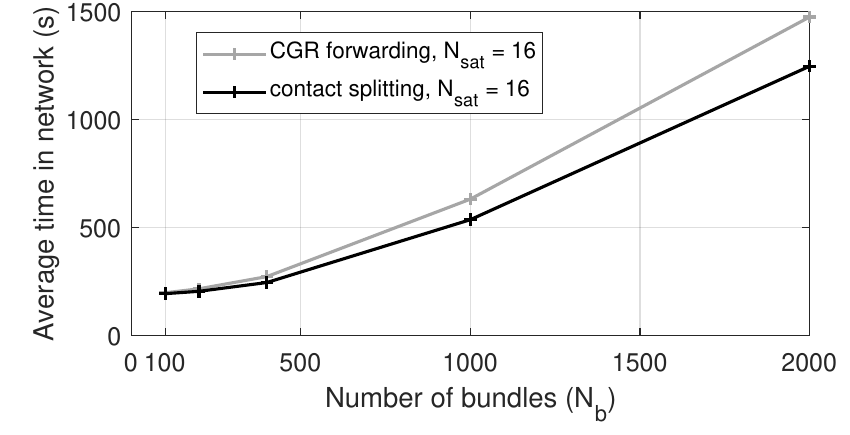}
	\caption{$N_\text{sat} = 16$.}
	\label{fig: timeInNet_16sats}
    \end{subfigure}
    \caption{Average time spent in the network ($R_j = 400$ bps).}
     \label{fig: timeInNet_const124}
  \end{figure}

Then, Figure \ref{fig: timeInNet_const124} illustrates the time spent in the network with $4$, $8$, and $16$ satellites (Constellations $1$, $2$, and $3$, respectively). The contact data rate is $400$ \gls{bps}. 
With four satellites (one satellite per orbital plane), the number of nodes is limited and they are more distant compared to constellations with higher numbers of satellites. Thus, the number of communication opportunities is limited, which constrains the routing degrees of freedom. In this case, the proposed algorithm shows a small improvement compared to the benchmark as seen in Figure \ref{fig: timeInNet_4sats}. When the number of satellites increases to $8$ or $16$, our algorithm shows greater improvements compared to the benchmark, particularly with high load ($N_\text{b}>1000$ in figures \ref{fig: timeInNet_8sats} and \ref{fig: timeInNet_16sats}).


\begin{figure}[!t]
	\centering
	\includegraphics[trim = 1.1in 3.5in 1.5in 3.5in, clip, width=\columnwidth]{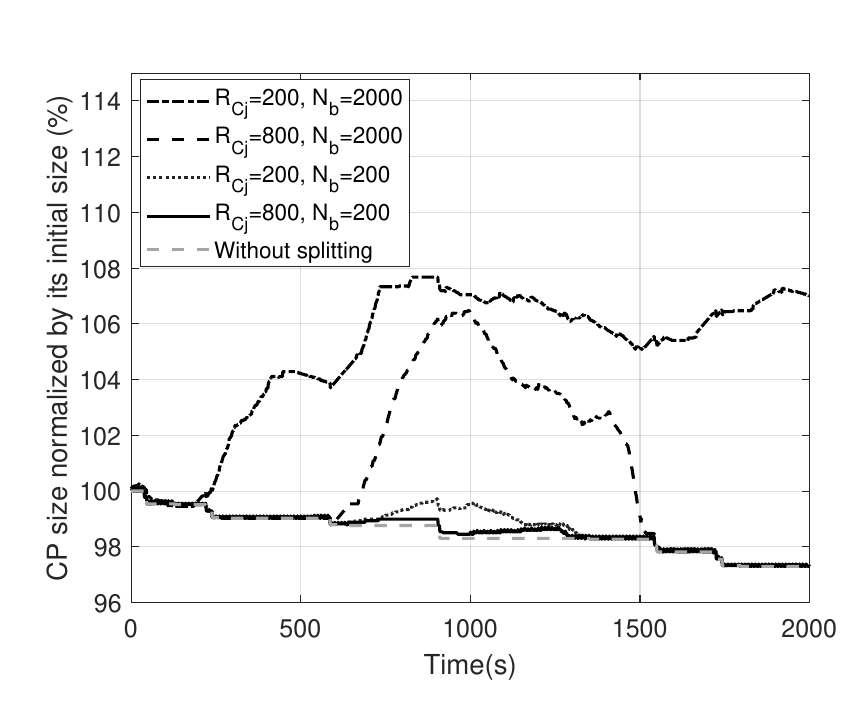}
	\caption{Impact of capacity management on the CP size, $N_\text{sat} = 16$.}
	\label{fig: capa_compl}
\end{figure}

Figure \ref{fig: capa_compl} illustrates the evolution of the \gls{cp} size over time, to assess the complexity induced by the proposed algorithm. 
The Y-axis shows the number of bundles in the \gls{cp} normalized by its initial size, which is of $3478$ contacts for Constellation $3$ (see Table \ref{tb: ConstChar}). 
The baseline curve, without capacity management, depicts the evolution of the \gls{cp} size over time as contacts expire and are removed from the \gls{cp}. Newly appearing contacts are not counted in this scenario. The other curves represent the evolution of the \gls{cp} size with our capacity management solution. 
In this case, splitting contacts can either increase the \gls{cp} size or leave it unchanged if contacts are shortened. 
Expired contacts are still removed from the \gls{cp}. 
We observe that our capacity management has a negligible impact on the \gls{cp} size and route search complexity under low load conditions. For instance, with $N_\text{b}=200$, the maximum difference between the baseline and our split \gls{cp} size is $1\%$ and $0.3\%$ for $R_j = 200$ and $800$ bps, respectively. Under high load conditions, the difference becomes more significant. 
For example, with $N_\text{b}=2000$, the increased frequency of bundle generation and routing forces the use of late contacts, as early contacts become fully booked. Consequently, the \gls{cp} size remains elevated until both split and original contacts expire. 
The highest difference with $N_\text{b}=2000$, observed for $R_j = 200$ at $t=1920$, is approximately $10\%$ compared to the baseline. 

\subsection{Impact of buffer management}
In this section, we evaluate the performance of the proposed buffer management solution. 
No contact capacity restriction is implemented, i.e., $R_j=\infty, \; \forall j$. 

\begin{figure}[!t]
    \begin{subfigure}[b]{0.5\textwidth}
     \includegraphics[trim = 1.1in 4.4in 1.5in 4.5in, clip, width=\columnwidth]{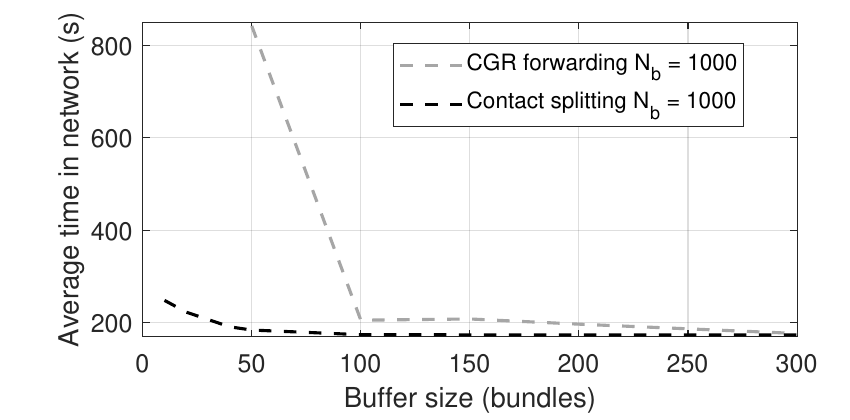}
	\caption{$N_\text{b} = 1000$.}
	\label{fig: timeInNet_Bvar1000}
    \end{subfigure}
    \begin{subfigure}[b]{0.5\textwidth}
      \includegraphics[trim = 1.1in 4.2in 1.5in 4.5in, clip, width=\columnwidth]{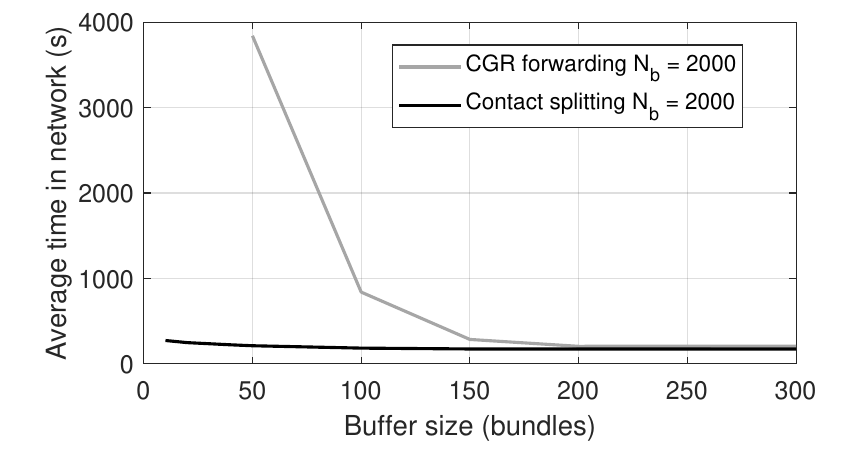}
	\caption{$N_\text{b} = 2000$.}
	\label{fig: timeInNet_Bvar2000}
    \end{subfigure}
    \caption{Average time spent in the network as function of the buffer size ($N_\text{sat}=16$).}
    \label{fig: timeInNet_Bvar}
  \end{figure}

The average time spent by a bundle in the network as a function of the buffer size per node, expressed by the number of bundles a buffer can store, is given in Figure \ref{fig: timeInNet_Bvar}. We consider $N_\text{sat}=16$ (constellation $3$). The buffer storage size is identical for all intermediate nodes. The source node has an infinite buffer size. The results are shown for $N_\text{b}=1000$ (Figure \ref{fig: timeInNet_Bvar1000}) and $N_\text{b}=2000$ (Figure \ref{fig: timeInNet_Bvar2000}) bundles generated over $2000$ seconds. As expected, when the storage capacity of the buffers increases, the time spent in the network is reduced. 
The gain is significant with the proposed solution, especially when the buffer sizes are limited.
To reach a performance comparable to infinite buffer capacity for $N_\text{b}=1000$, our solution requires buffers that can store $100$ bundles, to be compared with $200$ bundles to achieve this performance with the benchmark. 
The performance gap becomes even more pronounced with $N_\text{b} = 2000$: our solution achieves infinite-buffer performance with only $100$ bundles of buffer space, while the benchmark requires more than $300$ bundles to reach the same performance level.
Even with low buffer capacity, our buffer management performs comparably to an infinite buffer capacity case, unlike the benchmark. For example, with a buffer size of $50$ bundles, our buffer management results in a $17\%$ increase compared to an infinite buffer capacity, whereas the benchmark shows a $2000\%$ increase for $N_\text{b} = 2000$.


\begin{figure}[!t]
	\centering
\includegraphics[trim = 1.1in 3.5in 1.5in 3.5in, clip, width=\columnwidth]{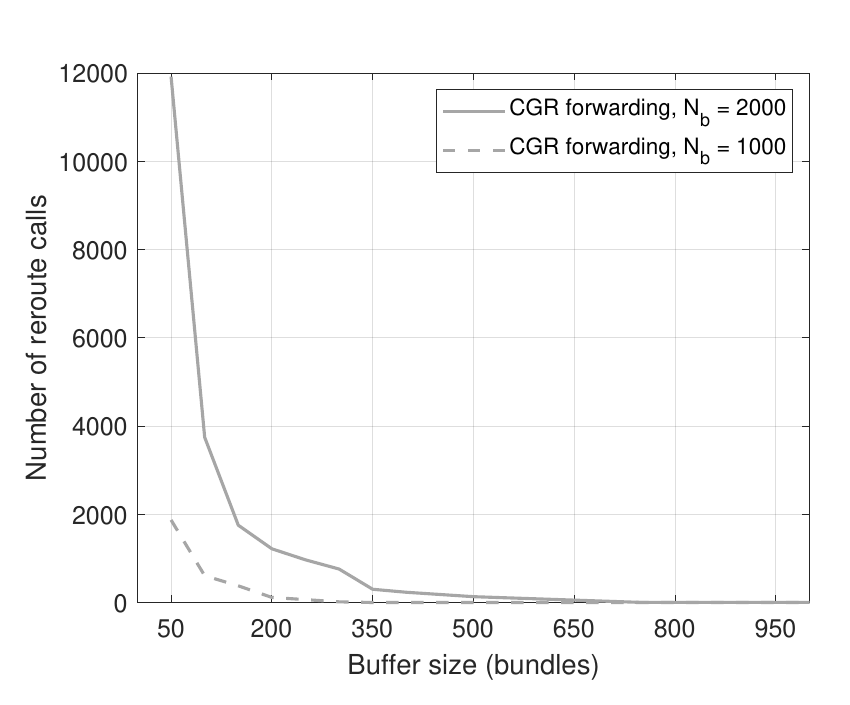}
	\caption{Number of reroute calls for CGR ($N_\text{sat} = 16$).}
	\label{fig: nRerouteBuffVar}
\end{figure}

Similarly to the capacity case, these differences are due to the number of rerouting events on intermediate nodes as shown in Figure \ref{fig: nRerouteBuffVar}, also for $N_\text{sat}=16$ (constellation $3$), when using \gls{sr}-\gls{cgr}. 
As expected, no rerouting had to be performed with the proposed solution.


\begin{figure}[!t]
	\centering
\includegraphics[trim = 1.4in 3.5in 1.5in 3.5in, clip, width=\columnwidth]{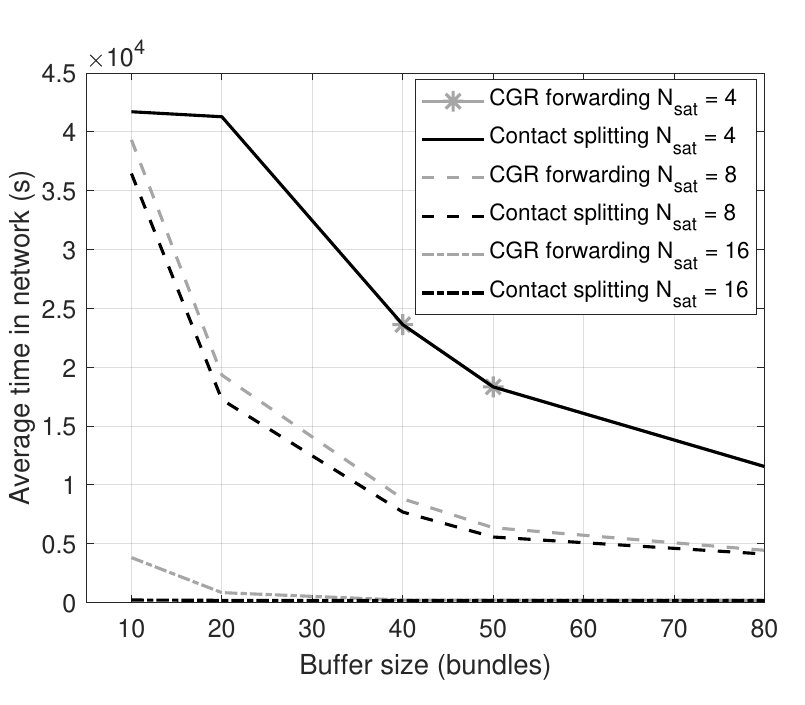}
	\caption{Average time spent in the network as function of buffer size for $N_\text{b} = 400$.}
	\label{fig: timeInNet_Constvar}
\end{figure}

Figure \ref{fig: timeInNet_Constvar} illustrates the performance impact of constellations with different sizes ($1$, $2$, and $3$) under buffer constraints, using $N_\text{b}=400$. In networks with fewer nodes, each node experiences higher utilization. With $N_\text{sat}=4$, the limited number of communication opportunities results in similar performance between both solutions, as the additional delay from limited storage capacity affects both approaches equally.
Increasing the number of satellites to $8$ highlights the effectiveness of our solution, which results in less time spent in the network than the benchmark.
Additionally, when the constellation is expanded to $16$ satellites, our proposed solution achieves very low delay, even with low buffer capacity per node.

\begin{figure}[!t]
	\centering
	\includegraphics[trim = 1.1in 3.5in 1.5in 3.5in, clip, width=\columnwidth]{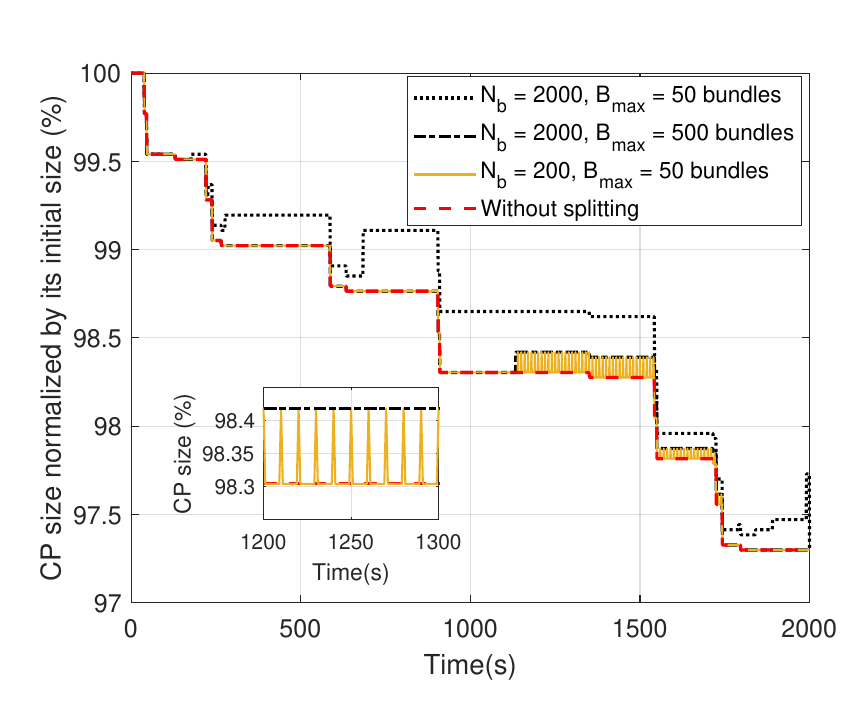}
	\caption{Impact of buffer management on the CP size.}
	\label{fig: buff_compl}
\end{figure}

In Figure \ref{fig: buff_compl}, we represent the evolution of the \gls{cp} size, normalized by its initial size ($3478$, see Table \ref{tb: ConstChar}), as function of time. The size of the \gls{cp} is reported each second. Without splitting for buffer management, the size of the \gls{cp} decreases with time as we only consider that the expired contacts are removed from the \gls{cp} as time advances. When considering buffer management, the splits are temporarily performed at bundle generation/routing time. With $N_\text{b} = 2000$, one bundle is generated each second. When the buffer size is $50$ bundles, the temporal splitting impacts the \gls{cp} size at almost all the time samples (also equal to the bundle generation time in this case). The impact of buffer management in this case is negligible, where the maximum is an increase of $0.43\%$ in the \gls{cp} size at $1992$ seconds. When the buffer size increases to $500$ bundles for the same $N_\text{b}$, the \gls{cp} size is not impacted most of the times due to high buffer capacity. It is nevertheless impacted between $1140$ and $1720$ seconds, where the maximum temporal increase is also negligible ($0.11\%$). We observe similar performance with $N_\text{b} = 200$ and a buffer size of $50$ bundles. Here, one bundle is generated and routed every $10$ seconds, causing a temporary increase in the \gls{cp} size, if it occurs, only during packet generation and routing. Outside of these times, the \gls{cp} size remains equal to its original size, without splitting.

\textcolor{blue}{\section{Future work: traffic-level extension and information sharing}
\label{sec:futurework}
A natural extension is to move from per-bundle routing to \emph{traffic-level} optimization with priorities. We maintain two priority-specific contact plans (CPs): High Priority (HP), non-preemptible, and Low Priority (LP), preemptible. Scheduling HP traffic consults the HP CP and writes the reservation into \emph{both} CPs. Any overlapping LP reservations are de-scheduled, their volumes returned to the LP CP, adjacent segments with $t_{\mathrm{end}}=t_{\mathrm{start}}$ are merged, and affected LP bundles are re-routed. Scheduling LP traffic consults and updates only the LP CP. Its reservations may later be invalidated by HP bookings.
Regarding information sharing in the network: the idea is then to keep the HP CP globally consistent, allow the LP CP to be useful yet preemptible, and bound control overhead while preserving the feasibility guarantees of the underlying CGR/SABR search. Other future directions (e.g., cislunar evaluation, distributed solutions, and the limitations listed above) are outlined in the conclusion (Section~\ref{sec: conclusions}).}

\section{Conclusion}
\label{sec: conclusions}

In this work, we proposed enhanced contact capacity and buffer management to improve the state-of-the-art \gls{cgr} routing algorithm. 
We proposed managing these resources by taking actions before or during the route search.
Our solution is a global solution: It considers and tracks the evolution of the contact capacity and buffer occupation in all contacts and nodes of the network. 
The resulting route is optimal and respects contact capacity and node buffer constraints. 

To manage contact capacity, we introduced contact splitting: \textcolor{blue}{it turns the initial contact set $V_c^0$ into the set $V_c$ of split contacts and enables tracking of contact portions previously consumed by routed bundles. For the buffer management, we proposed two techniques: temporary bundle-specific contact splitting (again modifying $V_c$) and edge pruning (removing from $E_c^0$ the edges that would cause overflow)}. These techniques help avoid any portion or succession of contacts that might result in buffer overflows \textcolor{blue}{as the route search runs on the feasible contact graph $(V_c,E_c)$}. These temporary modifications depend on the forecast buffer tables that track the future states of the node buffer occupancy levels as a function of time.

We compared the performance of the proposed solution with the benchmark \gls{sr}-\gls{cgr}. 
Our solution offers considerable gains in terms of time spent in the network and intermediate node complexity, measured as the number of rerouting calls. 
It is worth mentioning that the increase in \gls{cp} size due to our splitting actions, and thus the route search complexity, is relatively small. 

\textbf{Limitations.} Although the proposed solution offers improvements over the state-of-the-art while meeting capacity and buffer constraints, our solution remains centralized. 
\textcolor{blue}{Our evaluation assumes perfect contact-plan knowledge and no random disruptions. The impact of imperfect or delayed information is not assessed. We do not evaluate the algorithm under high error rates, random contact disruptions, or bursty traffic. The effectiveness under stochastic contact or buffer variations is left for future work. Quantitative evaluation of computational overhead for large-scale networks (e.g.\ many more satellites or longer time horizons) is also left for future work. We do not evaluate the overhead of global information exchange (e.g.\ forecast buffer tables, CP updates) or its feasibility in bandwidth-constrained deployments. Architectures such as Time-Sensitive Networking (TSN) may offer useful references for such analysis in future work. Validation on platforms such as the ONE Simulator or ION could be considered in future work to enhance reproducibility.}
The examination of an efficient distributed solution is therefore an interesting research direction for improvement of this work.


\textbf{Future work.} One could assess proposed cislunar constellations, from a communication perspective, using the presented routing technology.
We believe that the proposed routing algorithm may help with the design of the communication and storage systems to be embedded in the satellites of such lunar constellations.

One could also investigate traffic-level solutions rather than a per-bundle solution (see Section~\ref{sec:futurework} for a concrete extension with priority-based CPs and information sharing). 
In the state-of-the-art, like in this work, the route is chosen based on optimal criteria at the bundle level, not for the entire traffic to be transmitted. 
Therefore, this per-bundle strategy is unlikely to be optimal for the overall communication performance.
\textcolor{blue}{Evaluating network throughput (e.g., delivery throughput) could be considered in the scope of this future work.}
Finally, as mentioned in the previous ``limitations" subsection, investigating distributed solutions rather than a centralized one is also of interest. \textcolor{blue}{As noted in the limitations, future work may also include the directions listed there (ONE/ION validation, disruptions or imperfect CP, information-exchange overhead).}

\appendices
\section{Summary of Symbols}
\label{app:symbols}

Table~\ref{tb: notation} summarizes the mathematical notation.


\textcolor{blue}{%
\begin{table*}[t]
\color{blue}
\centering
\caption{Mathematical notation.}
\label{tb: notation}
\begin{tabular}{c>{\raggedright\arraybackslash}p{4.8cm}}
\hline
\textbf{Symbol} & \textbf{Meaning} \\
\hline
$\mathcal{N}$ & Set of nodes \\
$V_c^0$ (or CP), $V_c$ & $V_c^0$ = set of contacts in the contact plan (before splitting), $V_c$ = set of split contacts ($V$ is for vertices in the contact graph) \\
$\mathcal{P}_f(S,t_0)$ & Set of feasible routes for bundle of size $S$ and release time $t_0$ (paths in $(V_c^0,E_c^0)$ that are schedulable, contact-capacity feasible, and buffer feasible) \\
$E_c^0$, $E_c$ & Set of edges: $E_c^0$ = connectivity-only ($(i\to j)$ with $v_i=u_j$), $E_c$ = set of feasible edges (after pruning; $E_c\subseteq E_c^0$) \\
\hline
$\text{C}_i$ & Contact $\text{C}_i=(u_i,v_i,t_{1,i},t_{2,i},R_i,\delta_i)$ \\
$\text{C}_s$, $\text{C}_\text{end}$ & (CRP) Selected contact at current iteration; final contact of best route found (receiver = destination) \\
$t_{1,i}$, $t_{2,i}$ & Start and end time of contact $\text{C}_i$ \\
$t_{e_1}$, $t_{e_2}$ & Contact splitting: boundaries of the removed segment ($t_{1,j} \le t_{e_1} \le t_{e_2} \le t_{2,j}$; interval $(t_{e_1},t_{e_2})$ is erased) \\
$R_i$, $\delta_i$ & Data rate and one-way propagation delay of contact $\text{C}_i$ (bps) \\
$u_i$, $v_i$ & Sender and receiver nodes of contact $\text{C}_i$ \\
$s$, $d$ & Source and destination nodes of a bundle \\
$t_0$ & Bundle release time at source \\
\hline
$S_i$ & Bundle size (bits); $S_i$ = size of bundle $i$ \\
$b$ & Bundle being routed (buffer management, Algorithm~\ref{alg:bufferSteps}) \\
\hline
$P$, $P_f$ & Route (path) and feasible route; $P=(\text{C}_1,\ldots,\text{C}_k)$. Subscript $i_m$ = contact index along path (vertex $m$) \\
$P^*$ & Optimal route (minimum arrival time) \\
$t_{\text{Tx,}i}^{\text{C}_j}$ & Transmission time of bundle $i$ on contact $\text{C}_j$ \\
$\tau_i$ & Transmission start time on contact $\text{C}_i$ \\
$f_i$ & Arrival (finish) time at the receiver $v_i$ of contact $\text{C}_i$; $f_i=\tau_i+S/R_i+\delta_i$. Subscript $s$ for selected contact $\text{C}_s$ (CRP) \\
$A(P)$ & (Estimated) Arrival time at destination for route $P$ with a given contact graph \\
$W_i$ & Feasible start-time window for contact $\text{C}_i$: $W_i=[t_{1,i}, t_{2,i}-S/R_i]$ \\
\hline
\end{tabular}
\hspace{1.5em}
\begin{tabular}{c>{\raggedright\arraybackslash}p{4.8cm}}
\hline
\textbf{Symbol} & \textbf{Meaning} \\
\hline
$n$ & Node (generic index; e.g.\ in $B_n(t)$, $\bar B_n(t)$) \\
$I_{n}$ & Time intervals the bundle is stored in the buffer at node $n$\\
$B_n(t)$, $B_\text{max}$ & Buffer capacity at node $n$, time $t$; in this paper $B_n(t)=B_\text{max}$ (constant). $B_\text{max}$ = maximum buffer capacity (bits). \\
$Q_n(t)$ & Booked buffer load at node $n$, time $t$ (due to previously scheduled bundles) \\
$\bar B_n(t)$ & Residual buffer capacity at node $n$, time $t$: $\bar B_n(t)=\max\{0,B_n(t)-Q_n(t)\}$ \\
\hline
$\mathcal{Z}_\text{b}$ & Set of nodes with potential buffer overflow for bundle $b$ \\
$Z_k$ & Node with potential overflow ($Z_k \in \mathcal{Z}_\text{b}$) \\
$\mathcal{O}_{Z_k}$ & Set of potential overflows at node $Z_k$ \\
$t_{1,o_l}$, $t_{2,o_l}$ & Start and end time of overflow $o_l$ \\
$T_{1,s}$, $T_{2,s}$ & (CRP) Sets of overflow start and end times at the receiver of $\text{C}_s$ (used in buffer-aware route search) \\
\hline
$(V_c^0,E_c^0)$, $(V_c,E_c^0)$ & Connectivity-only contact graph (before and after splitting) \\
$(V_c,E_c)$ & Feasible contact graph (route search runs on this graph) \\
\hline
$N_\text{sat}$ & Number of satellites in the constellation \\
$N_\text{p}$ & Number of orbital planes \\
$N_\text{b}$ & Number of bundles (traffic load) \\
$\Delta\Omega$ & Angular separation between adjacent orbital planes \\
$k$ & Number of contacts in a path $P=(\text{C}_1,\ldots,\text{C}_k)$; or number of shortest routes in Yen's algorithm (implementation) \\
\hline
FEAP-CB & Feasible Earliest-Arrival Path with Capacity and Buffer constraints \\
\hline
\end{tabular}
\end{table*}}

\clearpage
\bibliographystyle{IEEEtran}
\bibliography{biblio.bib}

\end{document}